\documentclass[10pt,twocolumn]{article}
\usepackage[top=2cm, bottom=2cm, left=1.8cm, right=1.8cm]{geometry}

\usepackage{xspace}
\usepackage{times}
\usepackage{booktabs} 
\usepackage{subfigure}
\usepackage{graphicx}
\usepackage{soul}
\usepackage[keeplastbox]{flushend}
\usepackage{paralist}
\usepackage[square,sort&compress,comma,numbers]{natbib}
\usepackage{color}
\usepackage{url}
\usepackage[hang,flushmargin]{footmisc}

\usepackage{xcolor}
\usepackage{booktabs}
\usepackage{graphicx}
\usepackage{paralist}
\usepackage[small,bf]{caption}
\usepackage{subfigure}
\usepackage{times}
\renewcommand{\footnotesize}{\fontsize{8}{9}\selectfont}
\usepackage[compact]{titlesec}
\titlespacing*{\section}{0pt}{*3}{3pt}
\titlespacing{\subsection}{0pt}{*2}{2pt}
\titlespacing{\subsubsection}{0pt}{*2}{2pt}

\newcommand{\one}{({\em i}\/)\xspace}
\newcommand{\two}{({\em ii}\/)\xspace}
\newcommand{\three}{({\em iii}\/)\xspace}

\def\eg{\emph{e.g.,}\xspace}

\def\ie{\emph{i.e.,}\xspace}
\def\etal{\emph{et al.}\xspace}
\def\vs{\emph{vs.}\xspace}

\newcommand{\DW}{DW\xspace}

\renewenvironment{thebibliography}[1]{
  \begin{oldthebibliography}{#1}
    \setlength{\itemsep}{0.0em}
    \setlength{\parskip}{0.0em}
}
{
  \end{oldthebibliography}
}

\renewcommand{\footnoterule}{%
  \kern -3pt
  \hrule width 1in
  \kern 2pt
}

\makeatletter
\def\url@leostyle{%
  \@ifundefined{selectfont}{\def\UrlFont{}}%
  {\def\UrlFont{}}%
}
\makeatother
\usepackage[hyphenbreaks]{breakurl}

\definecolor{darkred}{RGB}{153,0,0}
\definecolor{darkblue}{RGB}{0,0,99}
\usepackage[colorlinks=true, linkcolor = darkred,   citecolor = darkred, urlcolor = darkblue]{hyperref}
\newif\ifcomment
\commentfalse

\ifcomment
\newcommand\gareth[1]{\textbf{\textcolor{blue}{GT: #1}}	}
\newcommand\ar[1]{\textbf{\textcolor{red}{AR: #1}}	}
\newcommand\sj[1]{\textbf{\textcolor{cyan}{SJ: #1}}	}
\newcommand\edc[1]{\textbf{\textcolor{red}{EDC: #1}}	}
\newcommand\ns[1]{\textbf{\textcolor{blue}{NS: #1}}	}
\else
\newcommand\gareth[1]{}
\newcommand\ar[1]{}
\newcommand\sj[1]{}
\newcommand\edc[1]{}
\newcommand{\ns}[1]{}
\fi

\usepackage{footnote}
\makesavenoteenv{tabular}

\newcommand{\paragraphbe}[1]{\vspace{0.75ex}\noindent{\bf #1}\hspace*{.3em}}

\begin{document}
\sloppy

\title{\bf Challenges in the Decentralised Web: The Mastodon Case\thanks{This paper appears in the Proceedings of 19th ACM Internet Measurement Conference (IMC 2019). Please cite the IMC version.}}
\date{}

\author{Aravindh Raman$^1$, Sagar Joglekar$^1$, Emiliano De Cristofaro$^{2,3}$, Nishanth Sastry$^1$, and Gareth Tyson$^{3,4}$\\[0.5ex]
\normalsize
$^1$King's College London, $^2$University College London, $^3$\hspace{-0.05cm}Alan Turing Institute, $^4$Queen Mary University of London\\
\small \{aravindh.raman,sagar.joglekar,nishanth.sastry\}@kcl.ac.uk, e.decristofaro@ucl.ac.uk, g.tyson@qmul.ac.uk }

\maketitle

\begin{abstract}

The Decentralised Web (\DW) has recently seen a renewed momentum, with a number of \DW platforms like Mastodon, PeerTube, and Hubzilla gaining increasing traction.
These offer alternatives to traditional social networks like Twitter, YouTube, and Facebook, by enabling the operation of web infrastructure and services without centralised ownership or control.
Although their services differ greatly, modern \DW platforms mostly rely on two key innovations: first, their open source software allows anybody to setup independent servers (``instances'') that people can sign-up to and use within a local community; and second, they build on top of federation protocols so that instances can mesh together, in a peer-to-peer fashion, to offer a globally integrated platform. 

In this paper, we present a measurement-driven exploration of these two innovations, using a popular \DW microblogging platform (Mastodon) as a case study. 
We focus on identifying key challenges that might disrupt continuing efforts to decentralise the web, and empirically highlight a number of properties that are creating natural pressures towards re-centralisation. 
Finally, our measurements shed light on the behaviour of both administrators (\ie people setting up instances) and regular users who sign-up to the platforms, also discussing a few techniques that may address some of the issues observed.

 \end{abstract}

\section{Introduction}

The ``Decentralised Web'' (\DW) is an evolving concept, which encompasses technologies broadly aimed at providing greater transparency, openness, and democracy on the web~\cite{wired2}. %
Today, well known social \DW platforms include Mastodon (a microblogging service), Diaspora (a social network), Hubzilla (a cyberlocker), and PeerTube (a video sharing platform).
Some of these services offer decentralised equivalents to web giants like Twitter and Facebook, mostly through the introduction of two key innovations. 

First, they decompose their service offerings into independent servers (``instances'') that anybody can easily bootstrap. 
In the simplest case, these instances allow users to register and interact with each other locally (\eg sharing videos), but they also allow cross-instance interaction via the second innovation, \ie \emph{federation}. This involves building on decentralised protocols to let instances interact and aggregate their users to offer a globally integrated service. 

\DW platforms intend to offer a number of benefits. For example, data is spread among many independent instances, thus possibly making privacy-intrusive data mining more difficult.
Data ownership is more transparent, and the lack of centralisation could make the overall system more robust against technical, legal or regulatory attacks. 

However, these properties may also bring inherent challenges that are difficult to avoid, particularly when considering the natural pressures towards centralisation in both social~\cite{wilson2009user,doerr2012rumors} and economic~\cite{stigler1958economies} systems.
For example, it is unclear how such systems can securely scale-up, how wide-area malicious activity might be detected (\eg spam bots), or how users can be protected from data loss during instance outages/failures.

As the largest and most popular \DW application~\cite{fedinfo, fediverse}, we choose \emph{Mastodon} as a relevant example to study some of these challenges in-the-wild. 
Mastodon is a decentralised microblogging platform, with features similar to Twitter. Anybody can setup an independent instance by installing the necessary software on a server. Once an instance has been created, users can sign up and begin posting ``{\em toots,}'' which are shared with followers.
Via federation, they can also follow accounts registered with other remote instances. %
Unlike traditional social networks, this creates an inter-domain (federated) model, not that dissimilar to the inter-domain model of the email system.
In this paper, we present a large-scale (case) study of the \DW aiming to understand the feasibility and challenges of running decentralised social web systems.
We use a 15-month dataset covering Mastodon's instance-level and user-level activity, covering 67 million toots. 
Our analysis is performed across two key axes: \one~We explore the deployment and nature of \emph{instances}, and how the uncoordinated nature of instance administrators drive system behaviours (Section~\ref{sec:char}); and \two~We measure how \emph{federation} impacts these properties, and introduces unstudied availability challenges~(Section~\ref{sec:availability}).
A common theme across our findings is the discovery of various pressures that drive greater centralisation; we therefore also explore techniques that could reduce this propensity. 

\paragraphbe{Main Findings.} Overall, our main findings include:

\begin{compactenum}

\item \emph{Mastodon enjoys active participation from both administrators and users}. There are a wide range of instance types, with tech and gaming communities being quite prominent. Certain topics (\eg journalism) are covered by many instances, yet have few users. In contrast, other topics (\eg adult material) have a small number of instances but a large number of users. 

\item \emph{There are {\bf\em user-driven} pressures towards centralisation}. Popularity in Mastodon is heavily skewed towards a few instances, driving implicit forms of centralisation. 10\% of instances host almost half of the users. This means that a small subset of administrators have a disproportionate impact on the federated system. 

\item \emph{There are {\bf\em infrastructure-driven} pressures towards centralisation}. Due to the simplicity and low costs, there is notable co-location of instances within a small set of hosting providers. We find that failures in these ASes can create a ripple effect that fragments the wider federated graph. 
For example, the Largest Connected Component (LCC) in the social follower graph reduces from 92\% of all users to 46\% by outages in five ASes. We observe 6 cases of these AS-wide outages within our measurement period. We also observe regular outages by individual instances (likely due to the voluntary nature of many administrators). Again, this has a notable impact: 11\% of instances are unavailable for half of our measurement period.

\item \emph{There are {\bf\em content-driven} pressures towards centralisation}. Due to the differing popularities of toots, we find that outages in just 10 instances could remove 62.69\% of global toots. To ameliorate this problem, we explore the potential of building toot replication schemes. For example, when a user has followers from different instances, the content posted by the user could be stored and indexed (persistently) in the followers' instances. By enabling this type of federation-based replication, availability improves so that only 11.4\% of toots are lost when the top 3 ASes are offline (rather than 70\% without).%
\end{compactenum}
\section{Mastodon}
\label{sec:primerMast}

In this section, we describe the basic operation of Mastodon, highlighting key terminology in bold. 
We refer readers interested in more details to a tutorial on The Verge~\cite{verge17}.

Mastodon is an open-source \DW server platform released in 2016~\cite{mastodon-git}. %
It offers microblogging functionality, allowing administrators to create their own independent Mastodon servers, aka \textbf{instances}. 
Each unique Mastodon instance works much like Twitter, allowing users to register new accounts and post \textbf{toots} to their followers. Users can also \textbf{boost} toots, which is the equivalent of retweeting in Twitter.

Instances can work in isolation, only allowing locally registered users to follow each other. 
However, Mastodon instances can also \textbf{federate}, whereby users registered on one instance can follow users registered on another instance.
This is mediated via the local instances of the two users. 
Hence, each Mastodon instance maintains a list of all remote accounts its users follow; this results in the instance \textbf{subscribing} to posts performed on the remote instance, such that they can be pulled and presented to local users.
For simplicity, we refer to users registered on the same instance as \textbf{local}, and users registered on different instances as \textbf{remote}.
Note that a user registered on their local instance does \emph{not} need to register with the remote instance to follow the remote user. Instead, a user just creates a single account with their local instance; when the user wants to follow a user on a  remote instance, the user's local instance performs the subscription on the user's behalf. 
This process is implemented using an underlying subscription protocol. To date, Mastodon supports two open protocols: oStatus~\cite{ostatus} and ActivityPub~\cite{activitypub} (starting from v1.6). This makes Mastodon compatible with other decentralised microblogging implementations (notably, Pleroma). 

When a user logs in to their local instance, they are presented with three timelines: \one~a \textit{home} timeline, with toots posted by the accounts whom the user follows; \two~a \textit{local} timeline, with all the toots generated within the same instance; and \three~a \textit{federated} timeline, with \emph{all} toots that have been retrieved from remote instances. 
The latter is not limited to remote toots that the user follows; rather, it is the union of remote toots retrieved by all users on the instance. 
The federated timeline is an innovation driven by Mastodon's decentralised nature; it allows users to observe and discover toots by remote users, and broaden their follower network.

\section{Datasets}
\label{sec:meth}

The goal of our paper is to better understand the nature of \emph{instances} and \emph{federation}, using Mastodon as a case study.
To do so, we rely on three primary datasets: \one~{\em Instances}: regular snapshots of instance metadata and availability; \two~{\em Toots}: historical user posts (toots) available on each instance; and \three~{\em Graphs}: the follower and federation graphs. 

\paragraphbe{Instances.}
We first extracted a global list of instance URLs and availability statistics from a dump provided by the \url{mnm.social} website. This site contains a comprehensive index of instances around the world, allowing us to compile a set of 4,328 unique instances (identified by their domain). These instances primarily run the Mastodon server software, although 3.1\% run the Pleroma software (\url{https://pleroma.social/}) instead. This is because, since 2017, these two implementations have federated together using the same front-end and federation protocol (ActivityPub). Hence, from a user's perspective, there is little difference between using Mastodon or Pleroma instances.

We obtain our data by using the monitoring service of \url{mnm.social}. Every five minutes, \url{mnm.social} connected to each instance's \url{<instance.name>/api/v1/instance} API endpoint.  The instance API returned the following information from each instance: name, version, number of toots, users, federated subscriptions, and user logins; whether registration is open and if the instance is online.

The data on \url{mnm.social} goes back to April 11, 2017. We collected all the data until July 27, 2018.
Maxmind was then used to map the IP address of each instance to their country and hosted Autonomous System (AS). Although some of these metadata items rarely change (\eg country), repeatedly fetching all the data every five minutes gives us fine-grained temporal data revealing how values evolved across time. %
Overall, we observe approximately half a billion data points.

\paragraphbe{Toots.} In May 2018, we crawled all available toots across the instances. To compile a list of instances to crawl, we started with the list collected via the \url{mnm.social} website. 
We then filtered these to only leave instances that were online during May 2018: this left 1.75K active instances which were accessible.
This obviously reveals a degree of churn in the instance population; across the 15 month measurement cycle 21.3\% of instance went offline and never came back online. 
We wrote a multi-threaded crawler to connect with each of these 1.75K instances, via their API, and collected their entire history of toots. 
To expedite the process, we parallelised this across 10 threads on 7 machines. 
Each thread was responsible for querying the federated timeline of its designated instance, iterating over the entire history of toots on the instance. 
To avoid overwhelming instances, we introduced artificial delays between API calls to limit any effects on the instance operations.
For each toot, the following data were collected: username, toot URL, creation date, media attachments, number of favorites, followers, and followings, toot contents, and hashtags.

Our toots dataset contains 67M public toots, generated by 239K unique users.
By comparing this against the publicly listed metadata obtained from the \emph{instance} dataset (recall this contained toot and user counts), we find that our dataset covers 62\% of the entire toot population.
The remaining 38\% of toots could not be collected: approximately 20\% of these toots were set to private, and the remainder were hosted on instances that blocked toot crawling. 

\paragraphbe{Follower and Federation Graphs.} We also crawled the followers and following lists for users in July 2018. To this end, we scraped the follower relationships for the 239K users we encountered who have tooted at least once. This was performed be iterating over all public users on each instance, and simply paging through their (HTML) follower list.\footnote{\url{https://<instance.name>/users/<user.name>/followers}}
This provided us with the ego networks for each user. 
We identified a user by their account name on a per-instance basis, as users must create one account per instance. 87.9\% of account names are unique to a single instance, while 8.3\% are on two instances and 3.8\% are on three or more. Note that it is impossible to infer if such accounts are owned by the same person and therefore we treat them as separate nodes. 

We then induced a graph, $G\big(V,E\big)$, in which user $V_i$ has a directed link to another user $V_j$ if $V_i$ follows $V_j$. This resulted in 853K user accounts, and 9.25M follower links. 
Conceptually, this graph is similar to the Twitter follower graph~\cite{kwak2010twitter}.
However, unlike Twitter, users can follow accounts on \emph{remote} instances, triggering the need for remote subscriptions between instances (\ie federation).
Hence, the creation of a link in $E$ can trigger an underlying federated subscription between the respective instances. 
We therefore created a second graph, $G_F\big(I,E\big)$, which consists of instance-to-instance subscriptions (hereafter instance \emph{federation graph}, for short).
$G_F\big(I,E\big)$ is induced by $G\big(V,E\big)$; a directed edge $E_{ab}$ exists between instances $I_a$ and $I_b$ if there is at least one user on $I_a$ who follows a user on $I_b$.

\paragraphbe{Twitter.} For comparisons, we also gathered data about outages in Twitter as well as its social network, dating back to when Twitter was at a similar age as Mastodon is now. 
For uptime statistics, we used~\url{pingdom.com}, a widely used website monitoring platform which probed Twitter every minute between February and December 2007 (when Twitter was roughly 1.5 years old). 
The data is obtained from the Internet Archive~\cite{archive}.
As a baseline social graph, we obtained the Twitter social graph from a 2011 dataset~\cite{leskovec2012learning}, which consists of users and their respective follower networks. 

\paragraphbe{Limitations.}
Our analysis reveals a highly dynamic system and, thus, our measurements are capturing a ``moving target.'' Similarly, we highlight that the timelines of each dataset differ slightly due to the progressive deployment of our measurement infrastructure. This, however, does not impact our subsequent analysis. 
It is also worth noting that we do not have comprehensive data on all toots, due to some instances preventing crawling (we have 62\% of toots). 
Thus, we have avoided performing deeper semantic analysis on toot content and, instead, focus on instance-level activities. 

\paragraphbe{Ethical Considerations.} The \emph{instances} dataset only contains public infrastructure information, whereas, the \emph{toots} dataset covers user information, and this might have privacy implications. 
However, we have exclusively collected public toot data, followed well established ethical procedures for social data, and obtained a waiver from the ethics committee at Queen Mary University of London.	 
All data is anonymised before usage, it is stored within a secure silo, and we have removed text content from our analysis.
Finally, note that our public dataset only covers infrastructure information and anonymised toot metadata.

\section{Exploring Instances}
\label{sec:char}

The first innovation that the \DW introduces is the distribution of services across independent \emph{instances}.
This means that these platforms emerge in a bottom-up fashion from the thousands of instances running their software. 
Hence, inspecting them provides a window into how bottom-up \DW decisions are made, and the implications they might have on the overall system. 
In this section, we perform a characterisation of the instances available in Mastodon, and how they are deployed.

\begin{figure}[t]
\centering
	\includegraphics[width=0.9\linewidth]{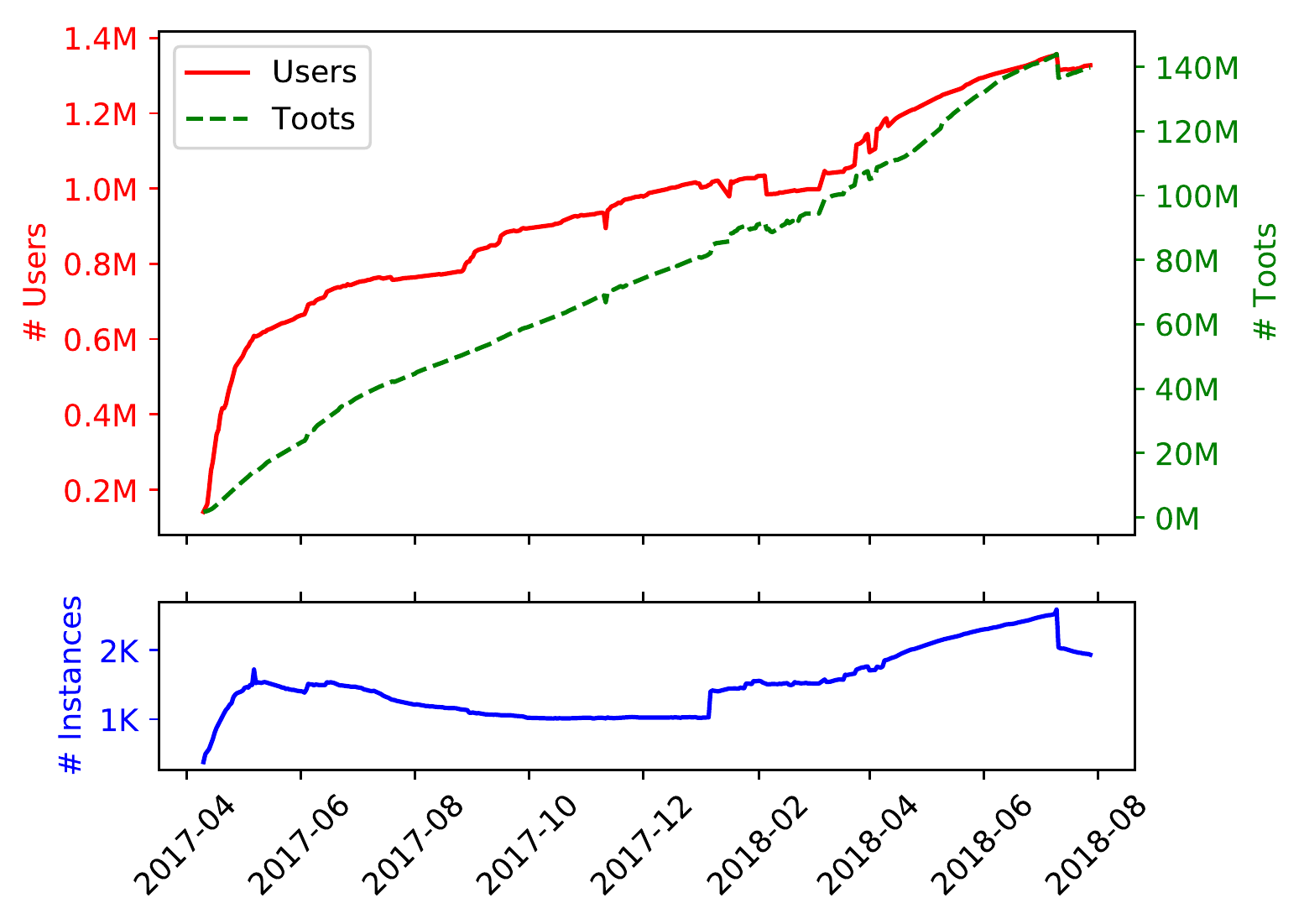}
	\vspace{-0.2cm}
	\caption{Available instances/user/toots over-time.}
	\label{fig:active_userstoots}
	\vspace{-0.2cm}	
\end{figure}

\subsection{Instance Popularity}

First, we quantify the growth and popularity of the 4,328 instances under study, and inspect how this relates to instance settings.

\paragraphbe{Instance Growth.}
We begin by briefly inspecting the recent growth of Mastodon.
Figure~\ref{fig:active_userstoots} plots the timeseries of the number of instances, users, and toots available per day.  
Mastodon is in a phase of growing popularity, with fluctuations driven by the arrival and departure of instances. 
Between April and June 2017, there was an increase in the number of instances (and users). 
However, while the number of instances reaches a plateau around July 2017 (only 6\% of the instances were setup between July and December 2017), the user population continues to grow during this period (by 22\%). 
Then, in the first half of 2018, new instances start to appear again (43\% growth).
We conjecture that this may have been driven by things like the \#DeleteFacebook campaign (popular at that time)~\cite{fortune18}, and sporadic bursts of media attention~\cite{wired2,verge17,what_is_mastodon}.
Closer inspection also reveals fluctuations in the instance population; this churn is driven by short periods of unavailability, where certain instances go offline (temporarily). 
We posit that this may have wider implications, and therefore deep dive into this in Section~\ref{sub:failures}.

\begin{figure*}[t]
\centering
	\subfigure[]{\includegraphics[width=0.3\linewidth]{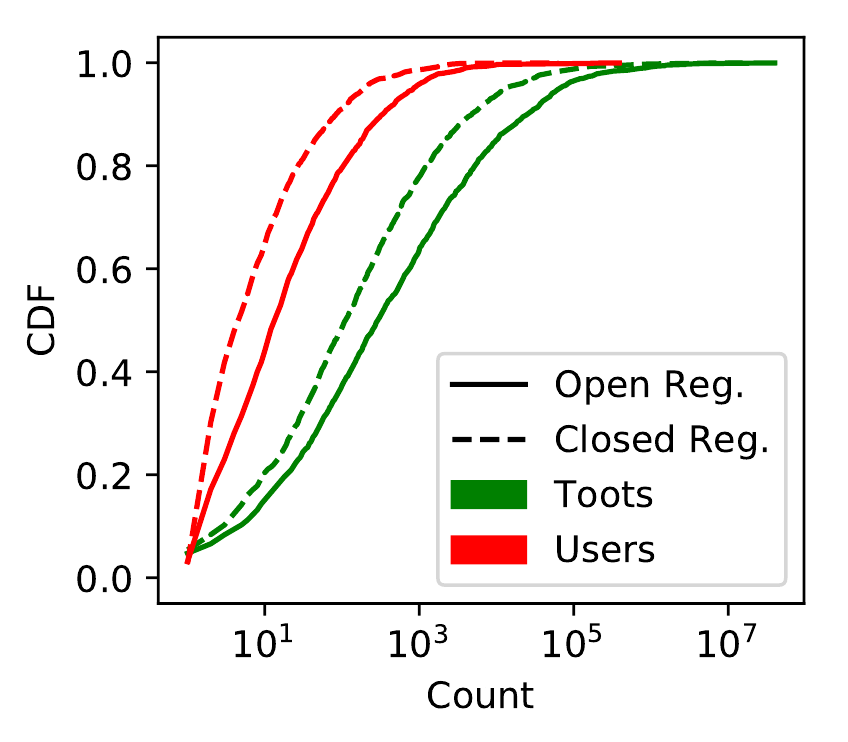} \label{fig:open_vs_closed:cdf}}
	\subfigure[]{\includegraphics[width=0.3\linewidth]{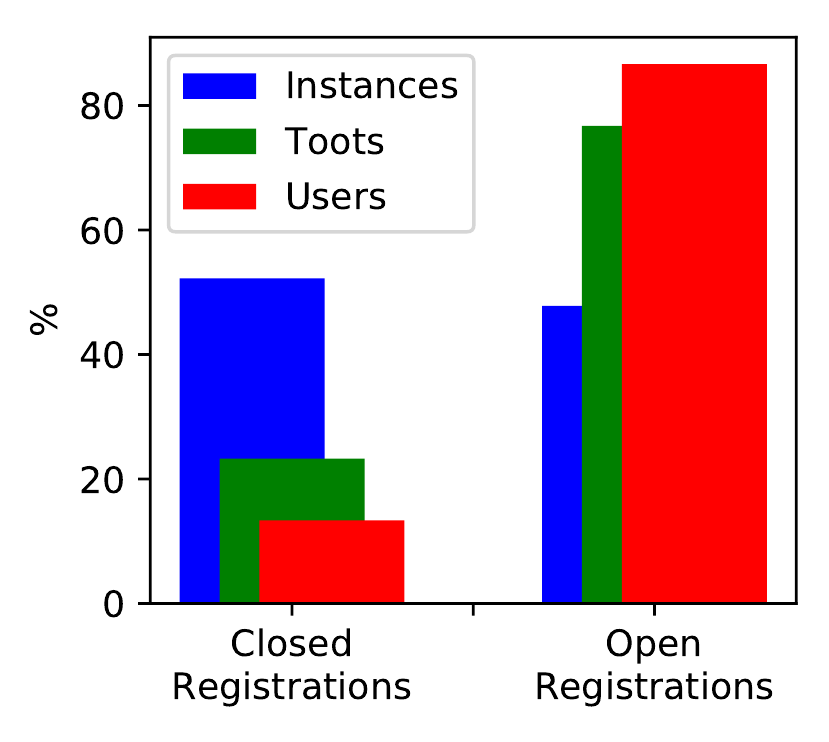} \label{fig:open_vs_closed:counts}}
	\subfigure[]{\includegraphics[width=0.3\linewidth]{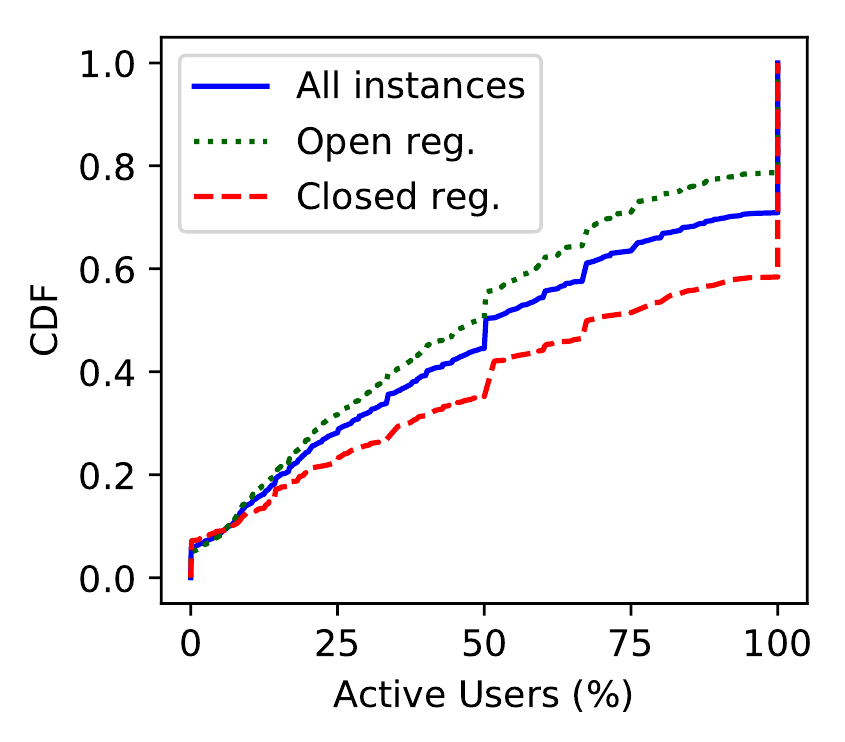} \label{fig:instanceActivity}}
	\vspace{-0.2cm}
	\caption{Dissecting Instances with open and closed (invite-only) registrations. $(a)$ Distribution of number of toots and users per-instance $(b)$ Number of instances, toots and users for open and closed registrations; $(c)$ Distribution of active users (max percentage of users logged-in in a week per instance) across all instances.}
	\label{fig:open_vs_closed}
	\vspace{-0.1cm}	
\end{figure*}

\paragraphbe{Open \vs Closed Instances.}
Mastodon instances can be separated into two groups: \one~\emph{open}, allowing anybody to register (47.8\% of instances); and \two~closed, requiring an explicit invitation from the administrator (52.2\%).
Figure~\ref{fig:open_vs_closed:cdf} presents a CDF of the number of users and toots per-instance, separated into open and closed instances. 
Overall, the user population distribution is highly skewed, with distinct traits of natural centralisation. For both kinds of instances (open and closed) the top 5\% of all instances have 90.6\% of all users. 
Similar patterns can be seen among toot generation, with the top 5\% of instances accumulating 94.8\% of toots. 
This of course confirms that, despite its decentralisation, Mastodon does not escape the power law trends observed in other social platforms~\cite{ugander2011anatomy,cha2010measuring,gilani2017bots,kwak2010twitter}.

Unsurprisingly, instances with an open registration process have substantially more users (mean 613 \vs 87). 
Inspecting user count alone, however, can be quite misleading, as open instances may accumulate more experimental (and disengaged) users. Figure~\ref{fig:open_vs_closed:counts} presents the breakdown of users, instances and toots across open and closed instances. Whereas the majority of users are associated with open instances, they create on average 94.8 toots per-person. 
In contrast, there are fewer users on closed instances, but they generate more toots per-capita (average of 186.65). To measure the activity level of users, each week, we compute the percentage of users who have actually logged into each instance (available in the~\emph{instance} dataset), and take the maximum percentage as the activity level of the instance.
Figure~\ref{fig:instanceActivity} presents the CDFs of the activity level per instance. This confirms that closed instances have more engaged populations: the median percentage of active users per closed instance is 75\%, compared to 50\% for active users in open instances. Despite this mix of instances, it is clear that a notable fraction of Mastodon users regularly login; for example, 13.73\% of users use Mastodon over once per-week).
This disparity in usage patterns, which is typical in any web system, does however increase the ``dominance'' of the instances that are associated with more active users. 

\subsection{Instance Categories}

\begin{figure}[t]
\centering
	\includegraphics[width=0.75\linewidth]{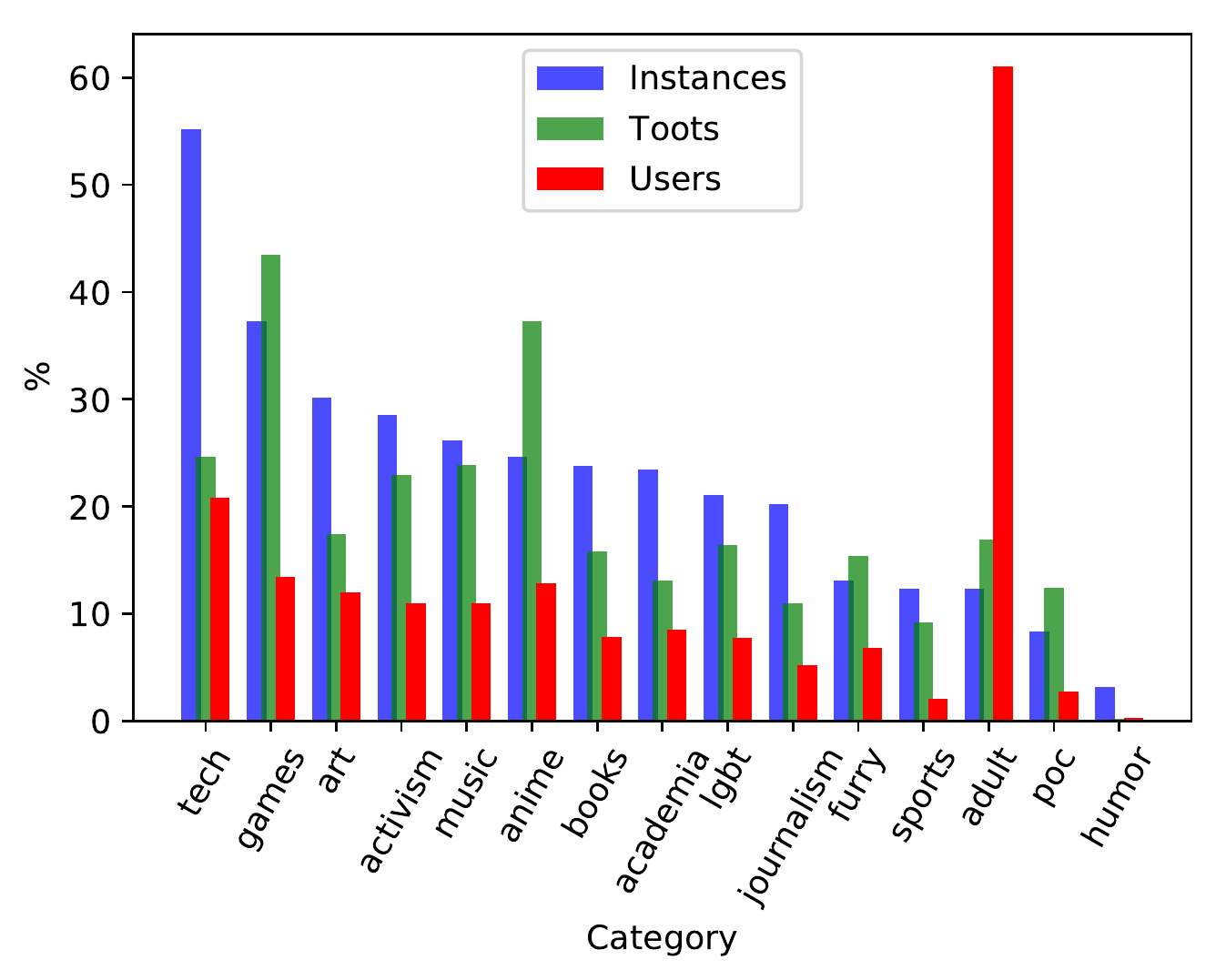}
	\vspace{-0.2cm}
	\caption{Distribution of number of instances, toots and users across various categories.}
	\label{fig:Instances}
		\vspace{-0.1cm}
\end{figure}

To improve visibility and/or to specialise in certain topics, instances can explicitly ``tag'' themselves with a category, as well as the activities they allow. Thus, these tags help in understanding the types of ongoing activities within Mastodon.

\paragraphbe{Category Tags.}
Overall, 697 out of the 4,328 instances report a self-declared category, taken from a controlled taxonomy of tags. Just 13.6\% of users and  14.4\% of toots are associated with these categorised instances.
51.7\% of these categories are labelled as ``generic''. Despite this relatively low coverage, it is still useful to inspect these tags to gain insight into ongoing usage. Note that the following statistics pertain to this subset of instances.
Figure \ref{fig:Instances} plots the distribution of instances, toots, and users across each category.
We identify 15 categories of instances. 
The majority of instances are categorised as tech (55.2\%), games (37.3\%) or art (30.15\%).
This is not entirely surprising, as Mastodon emerged from the tech community.

Figure \ref{fig:Instances}  also allows us to compare the popularity of each category when measured by number of instance \vs number of users, shedding light on the interests of the administrators \vs the general user population. 
Overall, the interests of these two groups coincide, albeit with some key discrepancies. 
For instance, while tech covers 55.2\% of instances, it only garners 20.8\% of users and 24.5\% of toots. By contrast, games corresponds to 37.3\% of instances, yet generate 43.43\% of all toots, suggesting they are highly active. 
Similar observations apply to Anime instances, where 24.6\% of instances contribute 37.23\% of global toots. 
There is, however, one clear outlier: adult instances constitute only 12.3\% of instances but attract 61.03\% of all users; 
that is, adult content is clearly a topic attracting a large user population, which is served by a small set of instances. 
This has been observed in other forms of online adult content, where websites with limited content sets gain large per-item interest~\cite{tyson2013demystifying}.
These contrasts suggest that Mastodon is a powerful tool for exploring and evaluating the differing behaviours of these communities.

\begin{figure}[t]
\centering
	\includegraphics[width=0.8\linewidth]{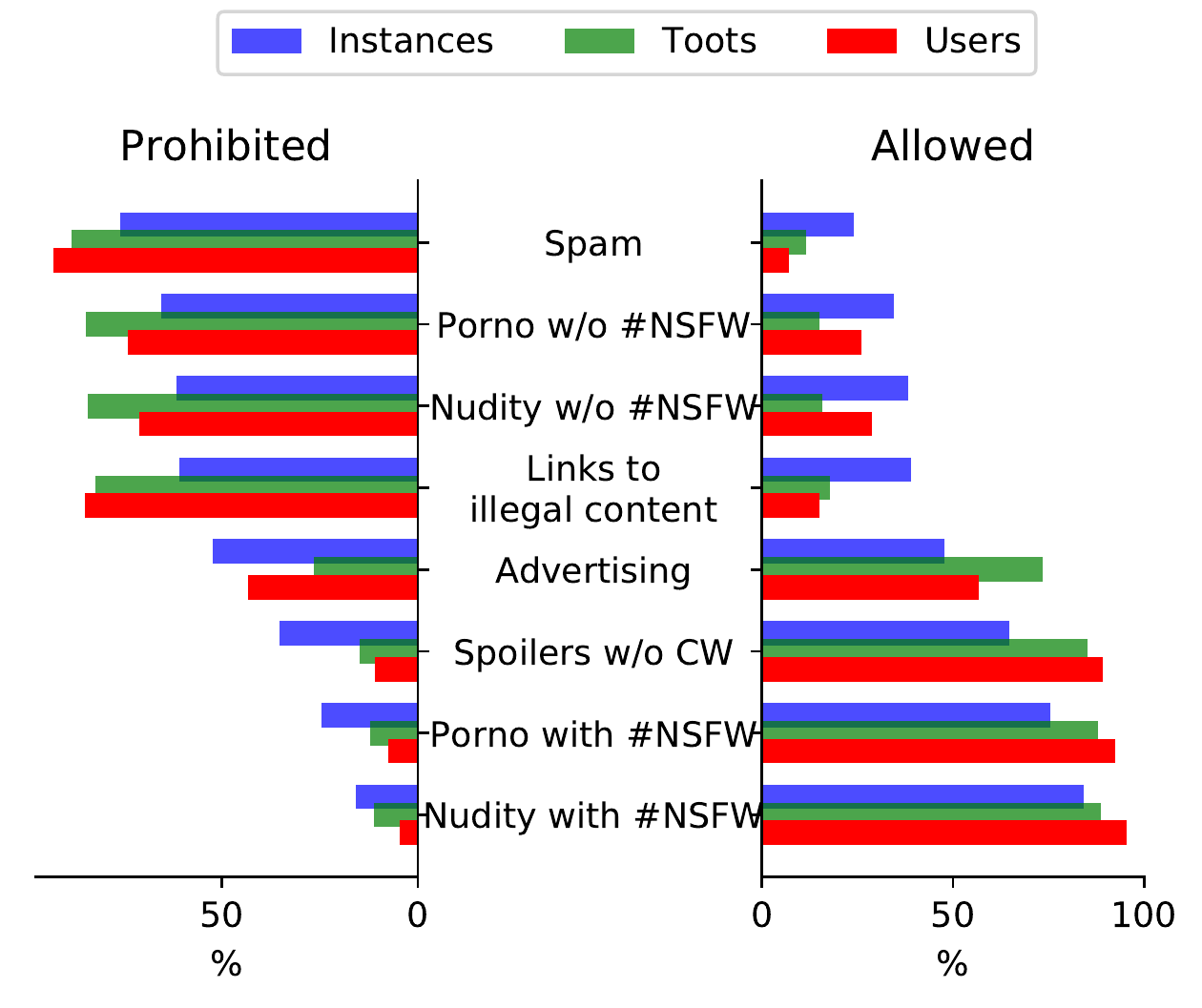}
	\vspace{-0.3cm}
	\caption{Distribution of instances and users across instances w.r.t.~prohibited and allowed categories.} 
	\label{fig:prohibitedAllowedDistrib}
\end{figure} 

\paragraphbe{Activity Tags.}
Due to its decentralised nature, it is also possible for individual administrators to stipulate what types of \emph{activity} are allowed on their instances.
This is necessary in a federated social network, as each instance can have different policy requirements. 
Out of the 697 instances, 17.5\% allow \emph{all} types of activity. The remaining instances explicitly specify a subset of activities acceptable on the instance. 82\% of these list at least one prohibited activity, whereas 93\% explicitly specify at least one acceptable activity.

Figure~\ref{fig:prohibitedAllowedDistrib} reports the number of instances, users, and toots associated with each activity category. The most regularly prohibited activity (left bar chart in the figure) is spam. 76\% of instances disallow it, followed by pornography (66\%), and nudity (62\%). 
These two activities are typically only prohibited when not tagged with \#NSFW (not safe for work): for instances that allow \#NSFW, the vast majority also allow nudity (84.3\%) and pornography (75.6\%).   
On the other hand, and quite curiously, some instances explicitly allow several of these activities (see right bar chart in Figure~\ref{fig:prohibitedAllowedDistrib}), \eg 24\% of instances allow spam. Unsurprisingly, these get few users: even though they make up 21\% of instances, they only hold 16\% of users. In contrast, instances allowing advertising have disproportionately large user groups (47\% of instances, but 61\% of users hosting and 75\% of the toots). 

Mastodon UI also has a ``content warning'' (CW) checkbox for the posters to give advance notice that there are spoilers in the content. Interestingly, while many instances prohibit posting spoilers without a content warning, many more explicitly allow this.
\subsection{Instance Hosting}
\label{sub:deployment}

\begin{figure}[!t]
	\centering
	\includegraphics[width=0.85\linewidth] 	{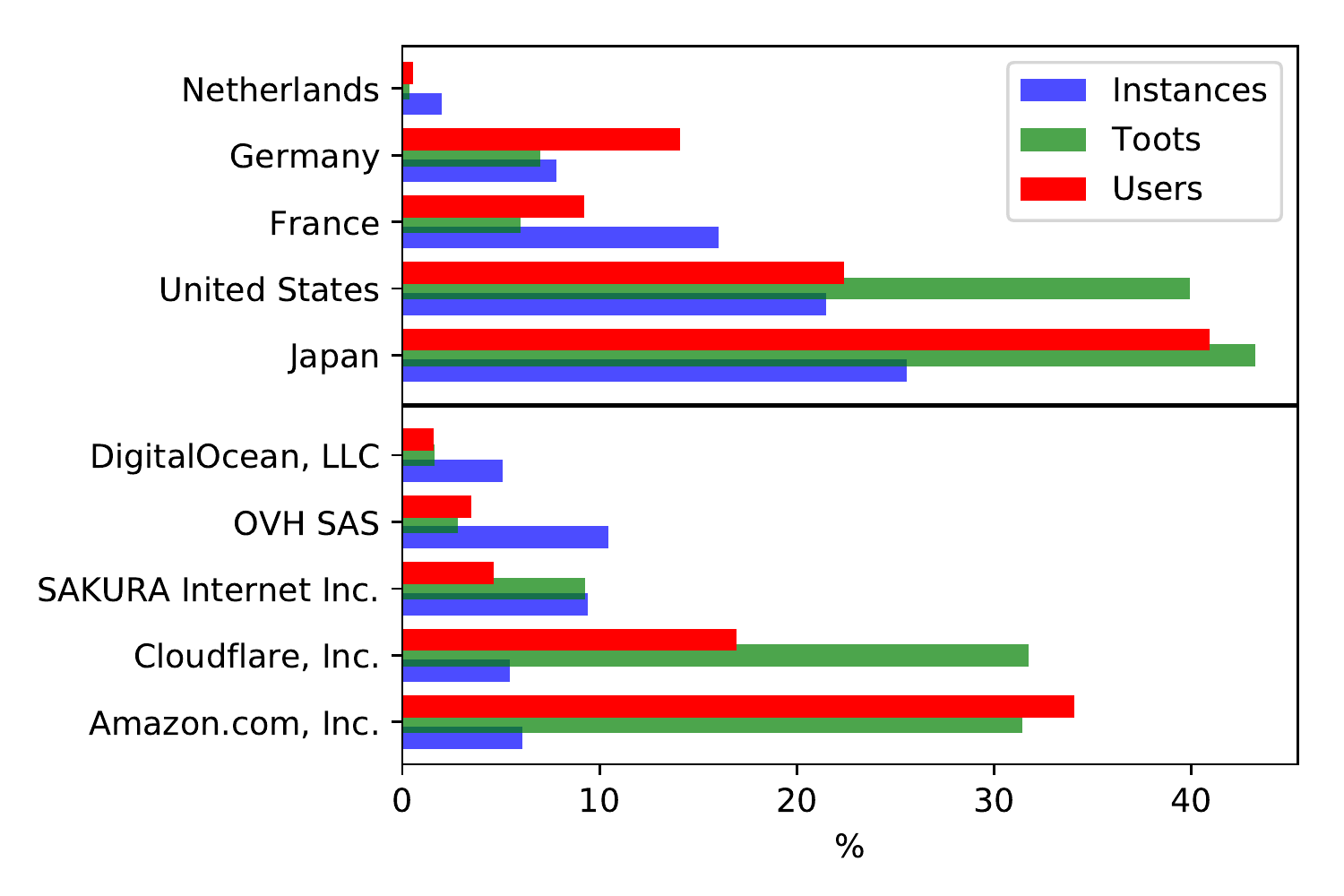}
	\vspace{-0.3cm}
	\caption{Distribution of instances, users, and toots across the top-5 countries (top) and  ASes (bottom).}
	\label{fig:instances_geo}
	\vspace{-0.15cm}
\end{figure}

\begin{figure}[!t]
	\centering
	\includegraphics[angle=0,width=0.9\linewidth]{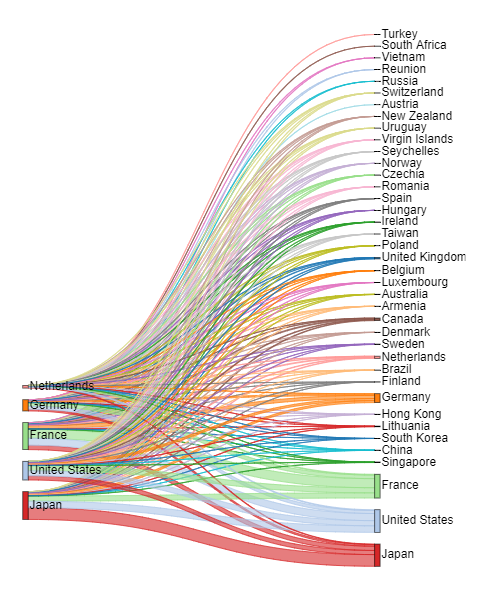}
	\vspace{-0.5cm}
	\caption{Distribution of federated subscription links between countries. The left axis lists the top-5 countries, and the lower access indicates the fraction of the federated links to instances in other countries.}
	\label{fig:peerInstances}
\end{figure}

Unlike a centrally administered deployment, where presence can be intelligently selected, Mastodon's infrastructure follows a bottom-up approach, where administrators independently decide where they place their instance. 
Figure~\ref{fig:instances_geo} presents a breakdown of the presence of instances, toots, and users across countries and Autonomous Systems (ASes). %

\paragraphbe{Countries.} Japan dominates in terms of the number of instances, users and toots. In total, it hosts 25.5\% of all instances, closely followed by the US which hosts 21.4\%. Closer inspection reveals that the ratio between the number of instances and number of users differ across countries though. For example, Japan hosts a just quarter of instances, yet gains 41\% of all users; in contrast, France hosts 16\% of instances, yet accumulates only 9.2\% of users. 
It is also worth noting that these countries are heavily interconnected, as instances must \emph{federate} together, \ie users on one instance may follow users on another instance (thereby creating federated subscription links between them, see Section~\ref{sec:primerMast}). 

To capture their interdependency, Figure~\ref{fig:peerInstances} presents a Sankey diagram; along the left axis are the top countries hosting instances, and the graph depicts the fraction of their federated subscriptions \emph{to} instances hosted in other countries (right axis). Unsurprisingly, the instances exhibit homophily: users of an instance follow other users on instances in the same country, \eg 32\% of federated links are with instances in the same country.
The top 5 countries attract 93.66\% of all federated subscription links. We posit that this dependency on a small number of countries may undermine the initial motivation for the \DW, as large volumes of data are situated within just a small number of jurisdictions; \eg 89.1\% of all toots reside on instances in Japan, the US, and France.

\paragraphbe{ASes.} 
Next, we inspect the distribution of instances across ASes; this is important as an over-dependence on a single AS, may raise questions related to data pooling or even system resilience. 
When looking at the ASes that host Mastodon servers (bottom of Figure~\ref{fig:instances_geo}), we observe instances spread across 351 ASes. On average, each AS therefore hosts 10 instances. 
This suggests a high degree of distribution, without an overt dependence on a single AS. 

That said, due to the varying popularity of these instances, the top three ASes account for almost two thirds (62\%) of all global users, with the largest one (Amazon) hosting more than 30\% of all users---even though it only is used by 6\% of instances. 
Cloudflare also plays a prominent role, with 31.7\% of toots across 5.4\% of instances. 
The reasons for this co-location are obvious: Administrators are naturally attracted to well known and low cost providers. 
Whereas, a centrally orchestrated deployment might replicate across multiple redundant ASes (as often seen with CDNs), this is more difficult in the \DW context because each instance is independently managed (without coordination). 
Although these infrastructures are robust, the failure (or change in policy) of a small set of ASes could therefore impact a significant fraction of users. Again, this highlights another form of tacit centralisation that emerges naturally within such deployments.

\subsection{Instance Availability }
\label{sub:failures}

We next explore the availability properties emerging from the bottom-up deployment.
Here, we define \emph{availability} as the ability to access and download its homepage.
We posit that the uncoordinated and (semi-)voluntary nature of some instance operations may result in unusual availability properties.

\begin{figure}[t]
	\centering
	\includegraphics[width=0.75\linewidth]{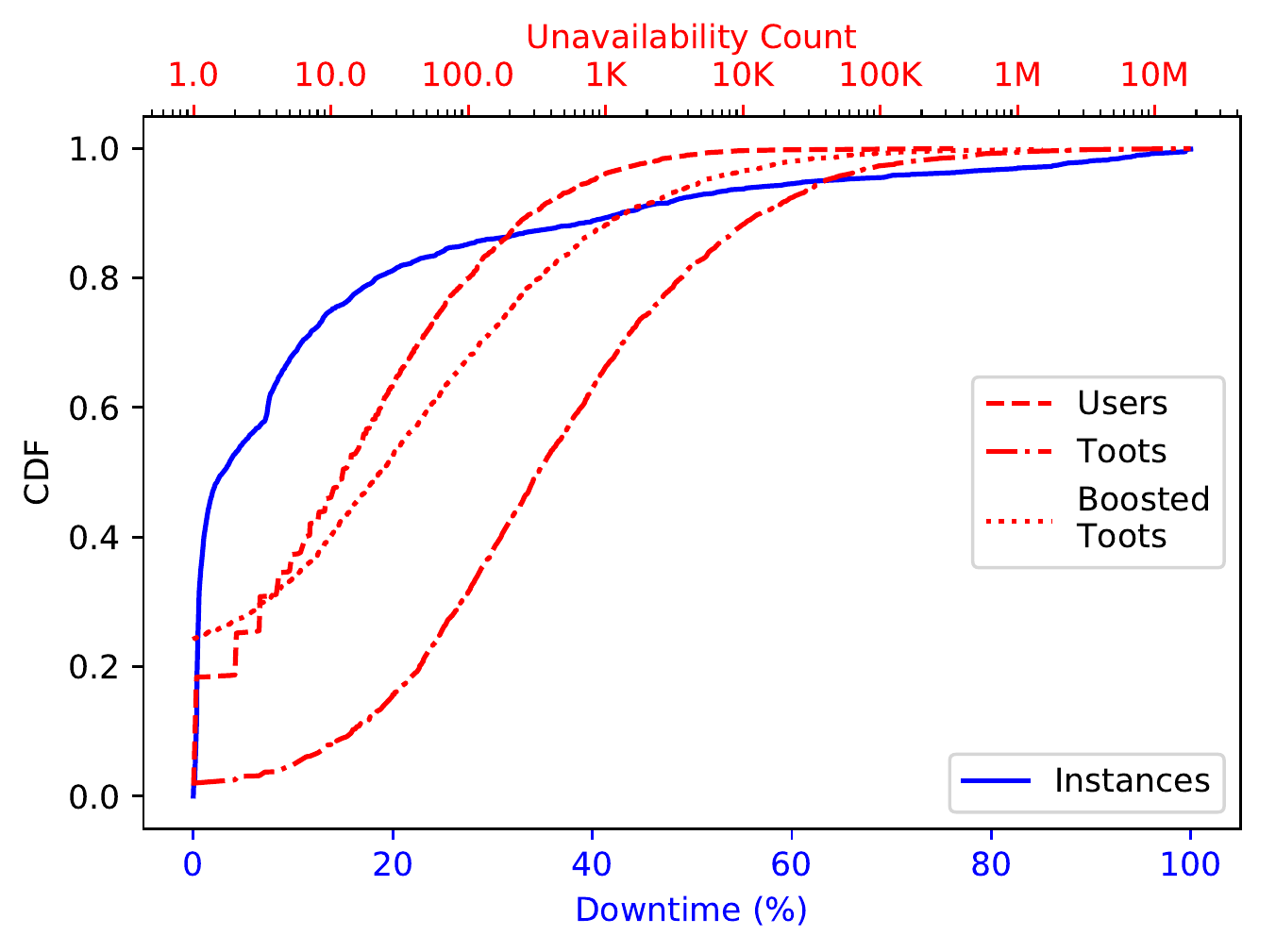}
	\vspace{-0.25cm}	
	\caption{CDF of the instance downtime (bottom x-axis) and distribution of unavailable users, toots, and boosted toots (top x-axis) when instances go down.  }
	\label{fig:instances_unavail}
	\vspace{-0.3cm}
\end{figure}

\paragraphbe{Instance Availability.}
We start by measuring the historical availability of our 4.3K instances over time (from the \emph{instance} dataset). We only count outages where an instance becomes unavailable, and then returns again within our 15 month measurement cycle (\ie we do not consider persistently failed instances as outages). 
Figure~\ref{fig:instances_unavail} plots the distribution of downtime of each instance over a 5-minute granularity (blue line, bottom X-axis). 
A sizeable portion of instances \textit{do} have relatively good availability properties -- about half of the instances have less than 5\% downtime. 
4.5\% of instances were even up for 99.5\% of the time (these popular instances covered 44\% of toots).
However, there is a long tail of extremely poor availabilities: 11\% of instances are inaccessible more than half of the time, which confirms that failures are relatively commonplace in Mastodon. This has obvious repercussions for both toot availability and the ability of users to access the platform. 

We cannot confirm the exact reasons for unavailability, nor can we state how reliable instances maintain their long uptime (although the operators might be employing load balancers and replicated back-ends). 
That said, it is likely that the voluntarily nature of many instance operators has a role, \ie instances with poor availability might simply be unpopular, and lack dedicated administrators. To test this, we count the number of toots and users that are unavailable during instance failures, see Figure~\ref{fig:instances_unavail} (red lines, top x-axis). 
Instances may go down multiple times, so we select the 95$^{th}$ percentile of these values for each instance.
Disproving our hypothesis, we find that failures happen on instances across the entire spectrum of popularity --- there are a number of  instances that host in excess of 100K toots which have outages. 

\begin{figure}[t]
\centering
	\includegraphics[width=0.8\linewidth]{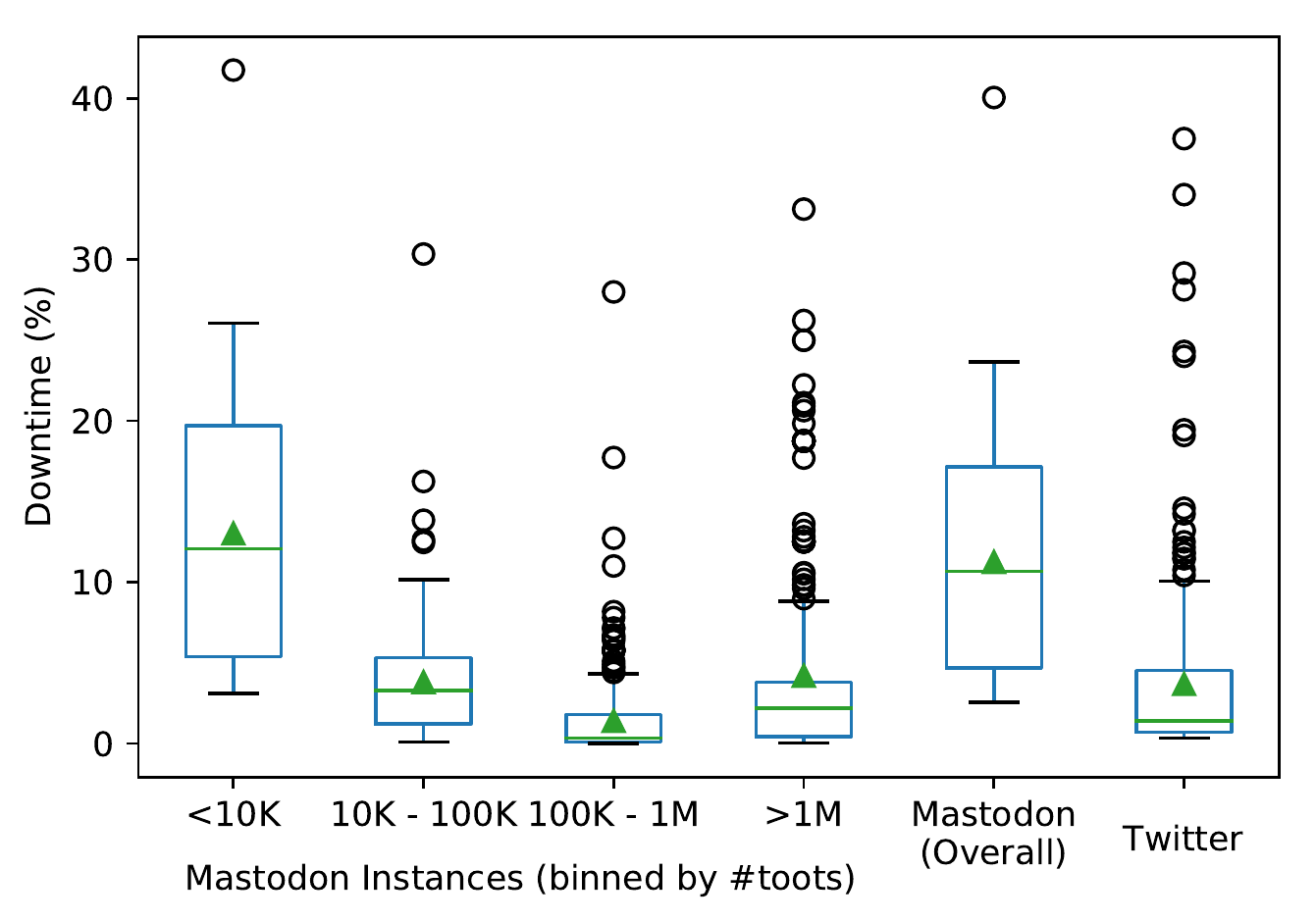}
	\vspace{-0.25cm}
	\caption{Distribution of per-day downtime (measured every five minutes) of Mastodon instances (binned by number of toots), and Twitter (Feb--Dec 2007).}
	\label{fig:twitDownTime}
		\vspace{-0.3cm}
\end{figure}

This is further explored in Figure~\ref{fig:twitDownTime}, which presents a box plot of daily downtime for Mastodon, where we separate instances based on their number of toots. 
Although small instances ($<$10K toots) clearly have the most downtime (median 12\%), those with over 1M toots actually have worse availability than instances with between 100K and 1M  (2.1\% \vs 0.34\% median downtime). In fact, the correlation between the number of toots on an instance and its downtime is -0.04, \ie instance popularity is not a good predictor of availability. 
The figure also includes Twitter's downtime in 2007 for comparison (see Section \ref{sec:meth}). 
Although we see a number of outliers, even Twitter, which was famous for its poor availability (the ``Fail Whale''~\cite{failWhale}), had better availability compared to Mastodon: its average downtime was just 1.25\% \vs 10.95\% for Mastodon instances.

\begin{figure}
	\centering
	\subfigure[]{\includegraphics[width=0.8\linewidth]{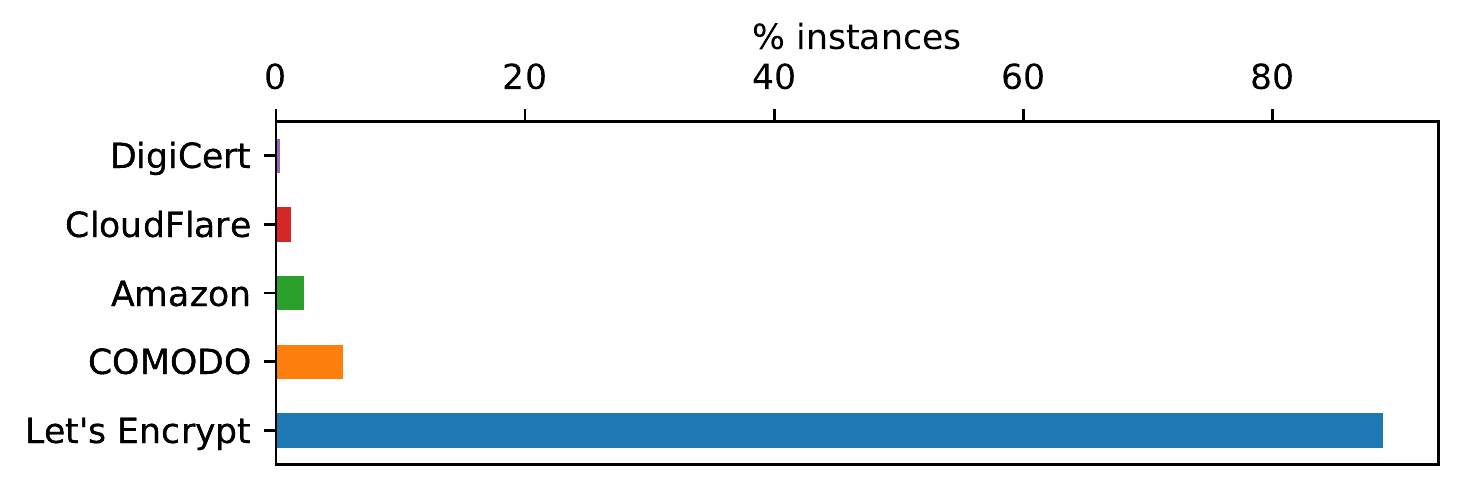}\label{fig:certFailures_a}}
	\hspace{0.2cm}
	\subfigure[]{\includegraphics[width=0.8\linewidth]{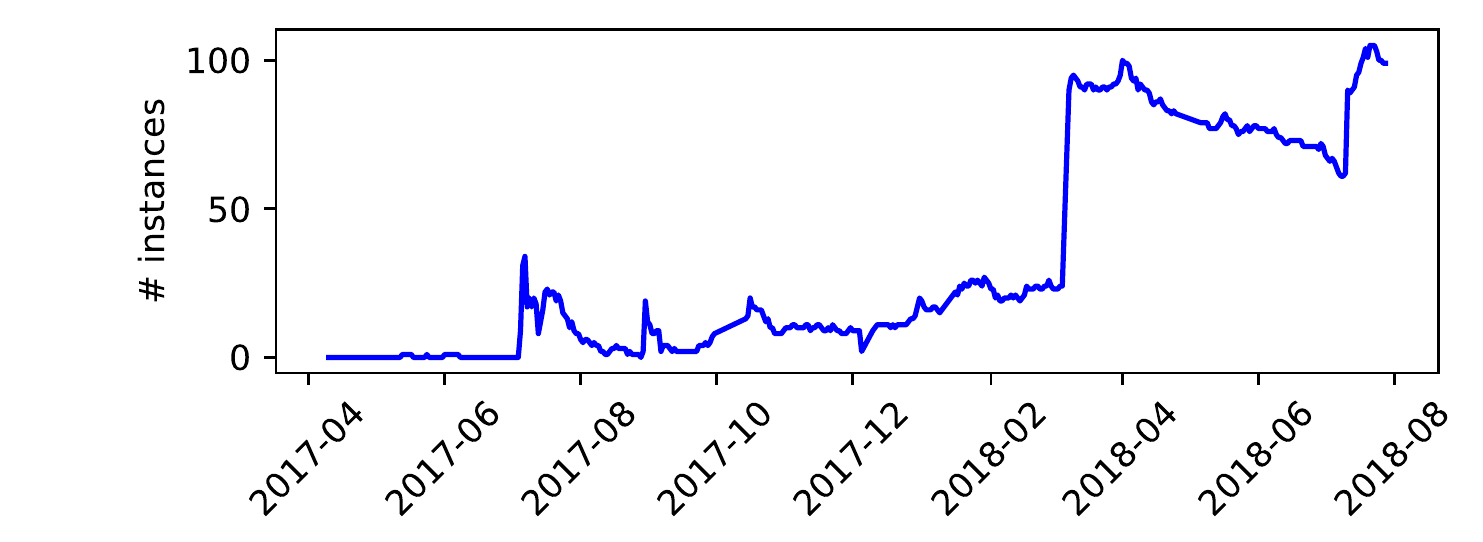}\label{fig:certFailures_b}}
	\vspace{-0.2cm}
	\caption{$(a)$ Footprint of certificate authorities among the instances.  $(b)$ Unavailability of instances (on a per-day basis).}
	\label{fig:certFailures} %
	\vspace*{-0.15cm}
\end{figure}

\paragraphbe{Certificate Dependencies.}
 Another possible reason for failures is third party dependencies, \eg TLS certificate problems (Mastodon uses HTTPS by default). 
To test if this may have caused issues, we take the certificate registration data from~\url{crt.sh}~\cite{crtsh}, and check which certificate authorities (CAs) are used by instances, presented in Figure~\ref{fig:certFailures_a}.
\emph{Let's Encrypt} has been chosen as CA for more than 85\% of the instances, likely because this service offers good automation and is free of cost~\cite{letsencrypt}. 
This, again, confirms a central dependency in the \DW.
We also observe that certificate expiry is a noticeable issue (perhaps due to non-committed administrators). 
Figure~\ref{fig:certFailures_b} presents the number of instances that have outages caused by the expiry of their certificates. 
In the worst case we find 105 instances to be down on one day (23 July 2018), removing nearly 200K toots from the system.
Closer inspection reveals that this was caused by the Let's Encrypt CA short expiry policy (90 days), which simultaneously expired certificates for all 105 instances. 
In total, these certificate expirations were responsible for 6.3\% of the outages observed in our dataset.

\paragraphbe{AS Dependencies.}
Another potential explanation for some instance unavailability is that AS-wide network outages might occur. 
Due to the co-location of instances within the same AS, this could obviously have a widespread impact.
To test this, we correlate the above instance unavailability to identify cases where \emph{all} instances in a given AS simultaneously fail --- this may indicate an AS outage. 
Table~\ref{tab:instance_failures_per_AS} presents a summary of the most frequent failures (we consider it to be an AS failure if  \emph{all} instances hosted in the same AS became unavailable simultaneously).
We only include ASes that host at least 8 instances (to avoid mistaking a small number of failures as an entire AS failure).
We  observe a small but notable set of outages. In total, 6 ASes suffer an outage. The largest is by AS9370 (Sakura, a Japanese hosting company), which lost 97 instances simultaneously, rendering 3.89M toots unavailable. The AS with most outages (15) is AS12322 (Free SAS), which removed 9 instances. 
These outages are responsible for less than 1\% of the failures observed, however, their impact is still significant. In total, these AS outages resulted in the (temporary) removal of 4.98M toots from the system, as well as 41.5K user accounts. 
Although this centralisation can result in such vulnerabilities, the decentralised management of Mastodon makes it difficult for administrators to coordinate placement to avoid these ``hot spots''.

\begin{table}[!t]
\footnotesize
\setlength{\tabcolsep}{3pt}
\begin{tabular}{l@{}rrrrrl@{}r@{}r}
\toprule
{\bf ASN} &   {\bf Instances} & {\bf Failures}  &  {\bf IPs} & {\bf Users} &   {\bf Toots} & {\bf Org.} & {\bf Rank} & {\bf Peers} \\
\midrule
AS9370 &   97 &  1 &        95 &  33.4K &  3.89M & Sakura & 2.0K & 10 \\
AS20473 &  22&   4 &       21 &   5.7K &   936K & Choopa & 143 & 150\\
AS8075 &  12 &   7 &      12 &   1.7K &    35.4K & Microsoft & 2.1K & 257\\
AS12322 &  9 &  15 &       9 &    123 &     4.4K & Free SAS & 3.2K & 63 \\
AS2516 &  9 &   4 &      8 &    559 &   102K & KDDI & 70 & 123 \\
AS9371 &  8&   14 &      8 &    165 &     4.7K & Sakura & 2.4K & 3\\
\bottomrule
\end{tabular}
\vspace{-0.3cm}
\caption{AS failures per number of hosted instances. Rank refers to CAIDA AS Rank, and Peers is the number of networks the AS peers~\cite{asRankCaida}.}
\label{tab:instance_failures_per_AS}
\vspace{-0.2cm}
\end{table}

\begin{figure}[t]
\centering
	\includegraphics[width=0.8\linewidth]{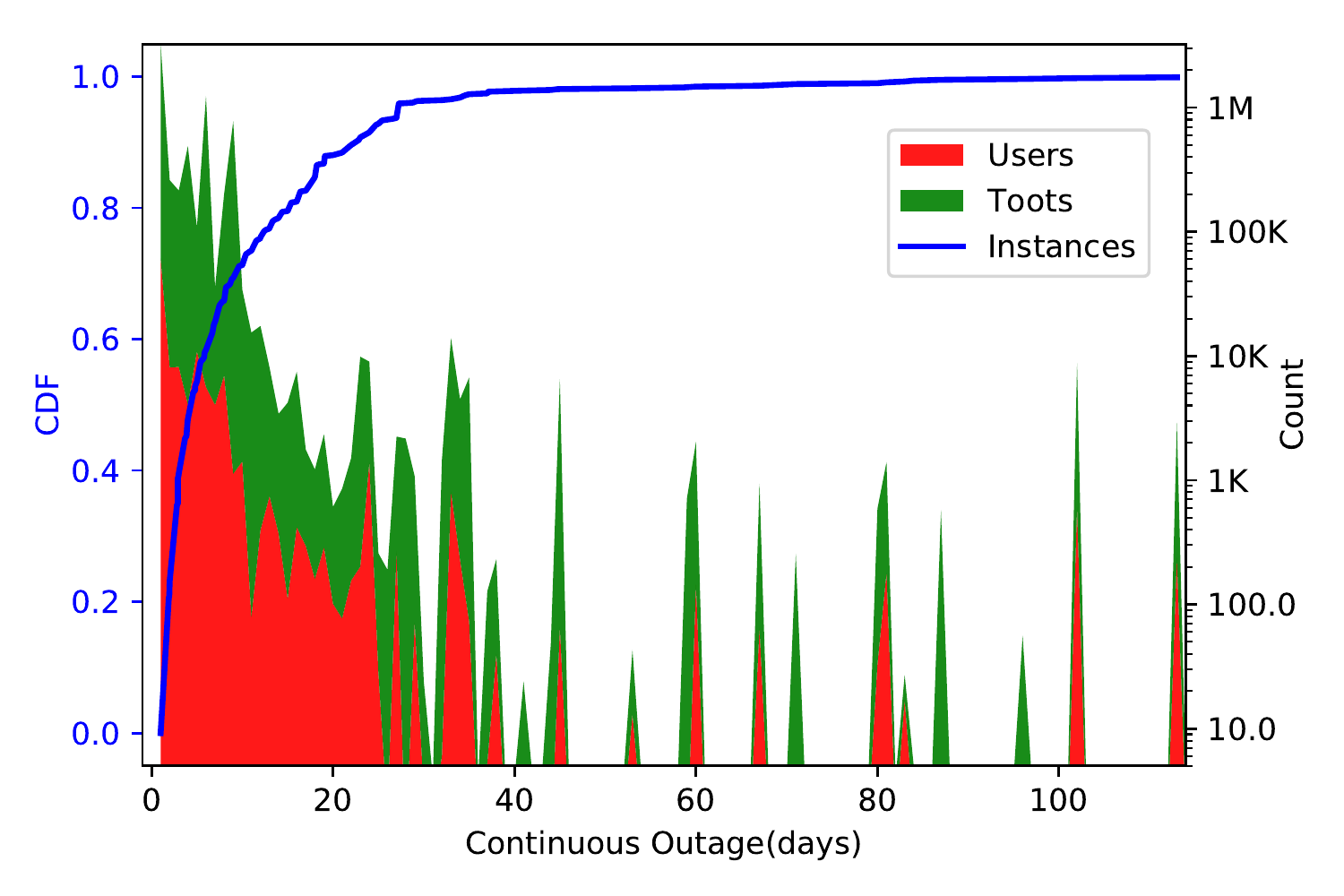}
	\vspace{-0.25cm}
	\caption{CDF of continuous outage (in days) of instances not accessible for at least one day (Y1-axis) and number of toots and users affected due to the outage (Y2-axis). }
	\label{fig:outageDays}
	\vspace{-0.15cm}	
\end{figure}

\paragraphbe{Outage durations.} Finally, for each outage, we briefly compute its duration and plot the CDF in Figure~\ref{fig:outageDays} (blue line, Y1-axis). 
While almost all instances (98\%) go down at least once, a quarter of them are unavailable for \emph{at least} one day before coming back online, ranging from 1 day (21\%) to over a month (7\%). 
Figure~\ref{fig:outageDays} also reports the number of toots and users affected by the outages:  14\% of users cannot access their instances for a whole day at least once.
Naturally, these measures are closely correlated to toot unavailability (\ie toots become unavailable when their host instance goes offline). 
In the worst case, we find one day (April 15, 2017) where 6\% of all (global) toots were unavailable for the whole day. 
These findings suggest a need for more reslient approaches to \DW management.

\begin{table*}
	\centering
	\footnotesize
	\begin{tabular}{lrrr@{ }rr@{ }rr@{ }rll}
		\toprule
		&  \multicolumn{1}{c}{\bf Toots from} & {\bf \#Home}  & \multicolumn{2}{c}{\bf Users} & \multicolumn{2}{c}{\bf Toots} & \multicolumn{2}{c}{\bf Instances}  & &\\
		{\bf Domain} &  {\bf Home Users} &  \multicolumn{1}{c}{\bf Users} & \multicolumn{1}{c}{\bf OD} & \multicolumn{1}{c}{\bf ID} & \multicolumn{1}{c}{\bf OD} & \multicolumn{1}{c}{\bf ID} & \multicolumn{1}{c}{\bf OD} & \multicolumn{1}{c}{\bf ID} & {\bf Run by} &  {\bf AS (Country)} \\
		\midrule
		mstdn.jp                  &      9.87M &       23.2K &    22.5K &  24.7K &  71.4M &  1.94M &      1352 &    1241 &   Individual & Cloudflare (US) \\
		friends.nico              &      6.54M &        8998 &     8809 &  23.3K &  37.4M &  2.57M &      1273 &    1287 &    Dwango & Amazon (JP)\\
		pawoo.net                 &      4.72M &       30.3K &    27.6K &  15.4K &  34.9M &  1.4M &      1162 &    1106 &    Pixiv & Amazon (US)\\
		mimumedon.com             &      3.29M &        1671 &      507 &   7510 &    435K &   366K &       420 &     524 &   Individual & Sakura (JP)\\
		imastodon.net             &      2.34M &        1237 &      772 &  10.8K &   2.37M &  1.52M &       711 &     865 &  Individuals (CF) & Amazon (US)\\
		mastodon.social           &      1.65M &       26.6K &    24.8K &  16.1K &  30.9M &   525K &      1442 &    1083 &    Individual (CF) & Online SAS (FR) \\
		mastodon.cloud            &      1.54M &        5375 &     5209 &    106 &   7.35M &      337 &      1198 &      39 &   Unknown & Cloudflare (US) \\
		mstdn-workers.com         &      1.35M &         610 &      576 &  12.5K &   4.18M &  1.98M &       735 &     850 &   Individual (CF) & Amazon (JP)\\
		vocalodon.net             &       914K &         672 &      653 &   8441 &   2.6M &   853K &       981 &     631 &   Bokaro bowl (A) & Sakura (JP) \\
		mstdn.osaka               &       803K &         710 &      363 &  1.64K &   2.68M &  2.1M &       561 &     862 &   Individual & Google (US) \\
		
		\bottomrule
	\end{tabular}
	\vspace{-0.2cm}
	\caption{Top 10 instances as per number of toots from the home timeline. (OD: Out Degree, ID: In Degree, CF: Crowd-funded, A: maintained by selling Bokaro Bowl album).}
	\label{tab:top_instances}
	\vspace{-0.3cm}
\end{table*}

\begin{figure}[t]
	\centering
		\includegraphics[width=0.8\linewidth] 	{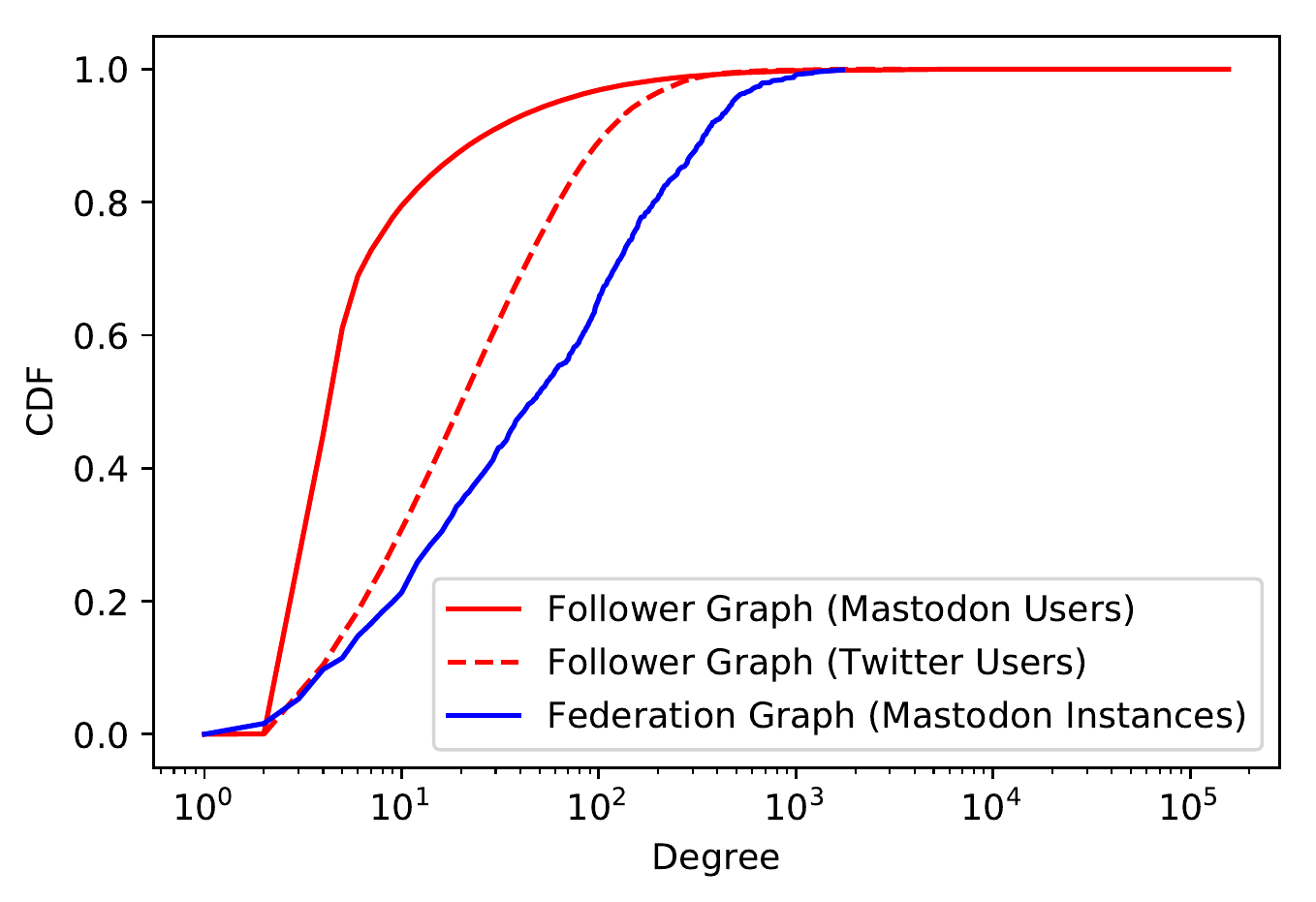}
\vspace{-0.3cm}
	\caption{CDF of the out-degree distribution of the social follower graph, federation graph, and Twitter follower graph.}
	\label{fig:degree_dist}
\vspace{-0.2cm}
\end{figure}

\section{Exploring Federation}
\label{sec:availability}

The previous section has explored the central role of independent \emph{instances} within the Mastodon ecosystem. 
The other major innovation introduced by the \DW is \emph{federation}. 
Here we inspect federation through two lenses: \one~the federated subscription graph that interconnects instances (Section~\ref{sub:social_failures}); 
and \two~the distributed placement and sharing of content (toots) via this graph (Section~\ref{sec:content}).
This section studies the resilience properties of \DW federation in light of the frequent failures observed earlier.

\subsection{Breaking the User Federation}
\label{sub:social_failures}
Federation allows users to create global follower links with users on other instances.
This means that instance outages (Section~\ref{sub:failures}) can create a transitive \emph{ripple effect}, \eg if three users on different instances follow each other, ${U_1 \rightarrow U_2 \rightarrow U_3}$, then the failure of the instance hosting $U_2$ would also disconnect $U_1$ and $U_3$ (assuming that no other paths exist). To highlight the risk, Figure~\ref{fig:degree_dist} presents the degree distribution of these graphs, alongside a snapshot of the Twitter follower graph (see Section~\ref{sec:meth}). We observe traditional power law distributions across all three graphs. 
Although natural, this creates clear points of centralisation, as outages within highly connected nodes will have a disproportionate impact on the overall graph structure~\cite{albert2000error}. 

To add context to these highly connected instances, Table~\ref{tab:top_instances} summarises the graph properties of the top 10 instances (ranked by the number of toots generated on their timeline).
As well as having very high degree within the graphs, we also note that these popular instances are operated by a mix of organisations, including companies (\eg Pixiv and Dwango), individuals, and crowd-funding. Ideally, important instances should have stable and predictable funding. 
Curiously, we find less conventional business models, \eg vocalodon.net, an instance dedicated to music that funds itself by creating compilation albums from user contributions. 

\paragraphbe{Impact of Removing Users.}
The above findings motivate us to explore the impact of removing nodes from these graphs.
Although we are primarily interested in infrastructure outages, we start by evaluating the impact of removing individual users from the social graph, $G\big(V,E\big)$. This would happen by users deleting their accounts. Such a failure is not unique to the \DW, and many past social networks have failed simply by users abandoning them~\cite{torok2017cascading}. 
Here, we repeat past methodologies to test the resilience of the
social graph by removing the top users and 
 computing two metrics: \one~the size of the Largest Connected
 Component (LCC), which represents the maximum potential number of
 users that toots can be propagated to (via shares); and \two~the
 number of disconnected components, which relates to the number of
 isolated communities retaining internal connectivity for propagating
 toots. These metrics have been used to characterise the 
 attack and error tolerance of social and other 
 graphs~\cite{albert2000error,yagan2013conjoining,jmir2018}.

\begin{figure}[t]
    \centering
    \includegraphics[width=0.8\linewidth]{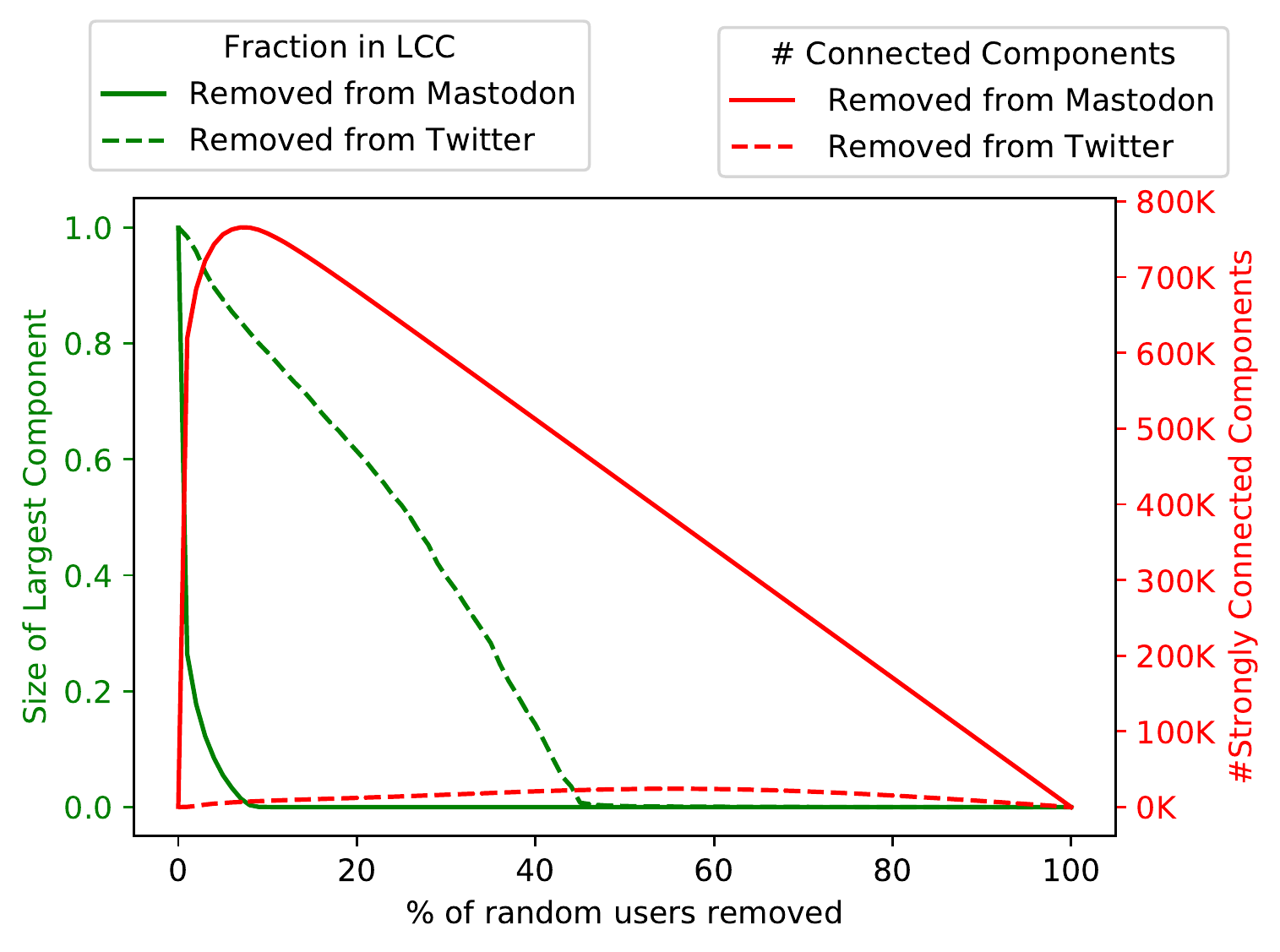}
	\vspace{-0.2cm}
    \caption{Impact of removing user accounts from
      $G\big(V,E\big)$. Each iteration (X axis) represents the removal
      of the remaining 1\% of the highest degree nodes.}
    \label{fig:sensitivity_users}
    	\vspace{-0.2cm}
\end{figure}

We proceed in rounds, removing the top 1\% of remaining nodes in each
iteration, and computing the size of the LCC in the remaining graph,
as well as  the number of new components created by the removal of
crucial connecting nodes. Figure~\ref{fig:sensitivity_users} presents
the results as a 
sensitivity graph. The results confirm that the user follower graph is
extremely sensitive to removing the highly connected users. 
Although Mastodon appears to be a strong social graph, with 99.95\% of
users in the LCC, removing just the top 1\% of accounts decreases the
LCC to 26.38\% of all users. 
As a comparison, we use the Twitter social graph from 2011 when Twitter was
a similar age as Mastodon is now (and beset with frequent ``fail
whale'' appearances~\cite{failWhale}).
Without any node removals, Twitter's LCC contained 95\% of users~\cite{cha2010measuring}; removing the top 10\% still leaves 80\% of users within
the LCC.  
This confirms that Mastodon's social graph, by comparison, is far more sensitive to
user removals. Although we expect that the top users on any platform will
be more engaged, and therefore less likely to abandon the platform, the \emph{availability} of top users to every
other user cannot be guaranteed since there is no central
provider and instance outages may remove these users. Indeed, individual instance failures, which we examine next,
can take out huge subsets of users from the global social graph.

\begin{figure}
    \centering
    \includegraphics[width=.95\linewidth]{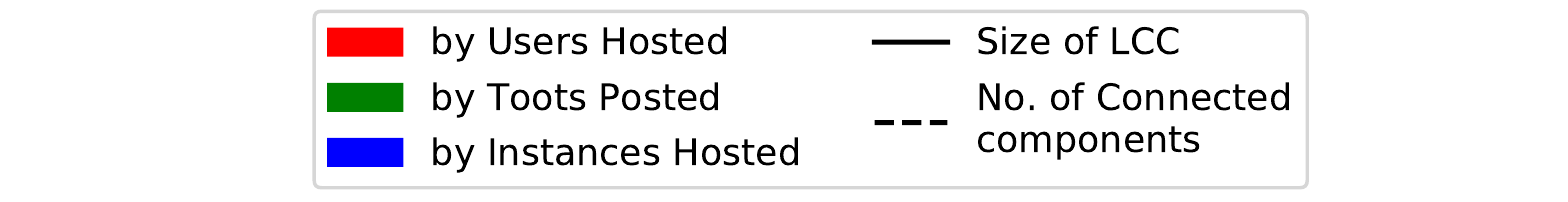}
        \subfigure[]{\includegraphics[width=0.495\linewidth]{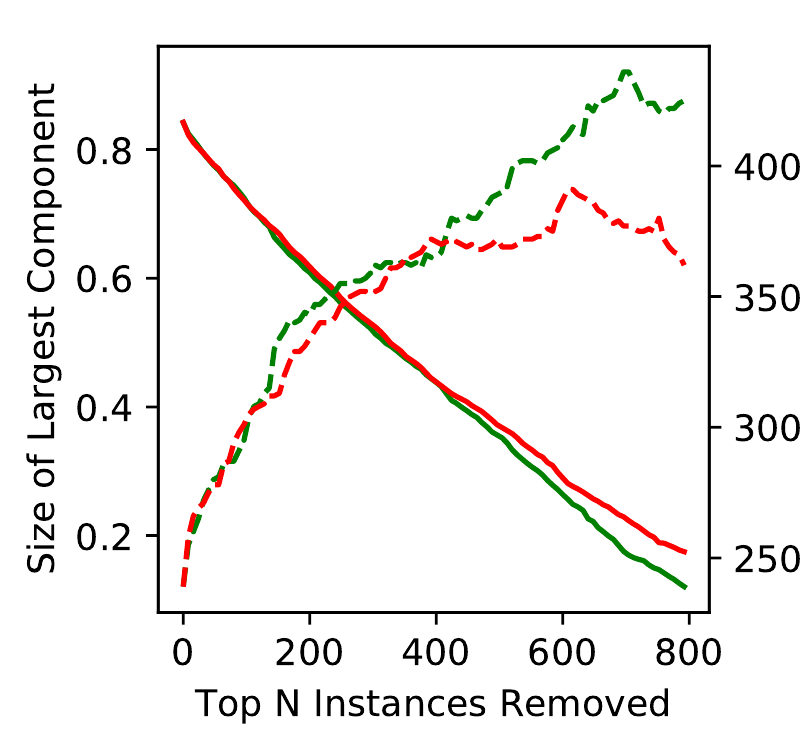}\label{fig:sens_instance}}
    \subfigure[]{\includegraphics[width=0.465\linewidth]{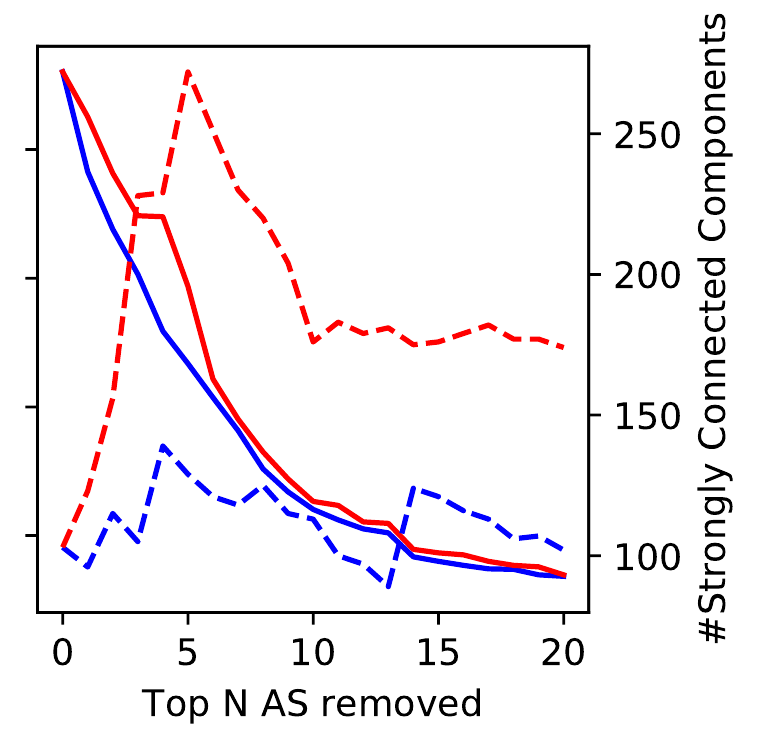}\label{fig:sens_AS}}
	\vspace{-0.2cm}
    \caption{Impact of node removal attacks. Each subfigure measures, on Y1 axis, LCC and, on Y2 axis, the number of components for $G_F\big(I,E\big)$, by removing: $(a)$ the top $N$ instances (each node in $G_F$ is an instance); and $(b)$ the top $N$ Autonomous Systems, including all instances hosted within.}
    \label{fig:sensitivityAll} %
    \vspace*{-0.25cm}
\end{figure}

\paragraphbe{Impact of Removing Instances.} 
As discussed earlier, instance failures are not uncommon, and can have
an impact that exceeds their local user base due to the (federated)
cross-instance interconnectivity of users in the social follower graph. Therefore, we next measure the
resilience of the instance federation graph ($G_F)$. %
In Figure~\ref{fig:sensitivityAll}(a), we report the impact of instance failures on $G_F$. We iteratively remove the top N instances, ordered by their size; we rank by both number of users (red) and number of toots (green). 
When ranking via either metric, we notice a remarkably robust linear decay in the size
of the LCC, and a corresponding increase in the number of components. 
Unlike the drastic breakdown of the social graph, this elegant degradation is caused by the more uniform degree
distribution of the federation graph (as compared against traditional social
networks~\cite{braunstein2016network}). We emphasise that the instance
federation graph shows the potential connectivity of  instances. However, individual
instance failures would still have an enormous impact on the
\emph{social} graph.

\paragraphbe{Impact of Removing ASes.} As discussed earlier, many instances are co-located in a small number of hosting ASes. We now inspect the impact of removing entire ASes, and thus all instances hosted within. 
Naturally, this is a far rarer occurrence than instance failures, yet they do occur (see Section~\ref{sub:failures}). We do not present this as a regular situation, but one that represents the most damaging theoretical impact. For context, AS-wide collapse might be caused by catastrophic failures within the AS itself~\cite{lakhina2004characterization,gunawi2016does} or via their network interconnections~\cite{giotsas2017detecting}.

Figure~\ref{fig:sensitivityAll}(b) presents the LCC and number of
components for $G_F$, while iteratively removing the
ASes, ranked by both the number of instances (blue) and number of
users (red).
At first, we see that 92\% of all instances are within a single 
LCC. This LCC covers 96\% of all users.
The graph shows that removing large ASes, measured by the number of instances
(blue), has a significant impact on $G_F$. The size of the
largest connected component decreases similarly whether we remove the
largest ASes when ranked by instances hosted (blue) or by number of
users (red). However, the number of connected components in $G_F$ increases
drastically when we remove the ASes hosting the largest users rather
than ASes ranked by number of instances: the removal of just 5 ASes
shatters the federation graph into 272 components when sorted by users
hosted, compared to just  139 when ranking by the \#instances in the
AS. This is explained by the central role of a few ASes: the top 5
ASes by users cover only 20\% of instances (yet comprise 85\% of
users); when ranked by number of instances, the top 5 covers 42\% of
instances (and 33.6\% of users). 
Thus, when AS failures occur, Mastodon shows
significantly worse resilience properties than previously seen for
just instance failures (Figure~\ref{fig:sensitivityAll}(a)). This is
driven by the fact that the top five ASes by number of instances
hosted --- OVH SAS (FR), Scaleway (FR), Sakura Internet (JP), Hetzner
Online (DE), and GMO Internet (JP) 
--- account for 42\% of all instances.  Their removal yields a 49\%
reduction in the size of LCC in the federation graph, leaving behind
an LCC which only covers 45\% of instances and 66\% of users.  
This constitutes a radical drop in the capacity of Mastodon to
disseminate toots via the federated subscription links. Indeed,
removing them not only wipes out a large number of nodes, but also
results in a smaller number of components which still remain. That
said, the linear degradation of the instance
federation graph discussed previously provides some limited protection
against a more catastrophic failure as observed with the Mastodon
social graph.
Although a rare occurrence, we therefore argue that techniques to avoid overt dependency on individual hosting ASes would be desirable.

\begin{figure}[t]
\centering
	\includegraphics[width=0.8\linewidth]{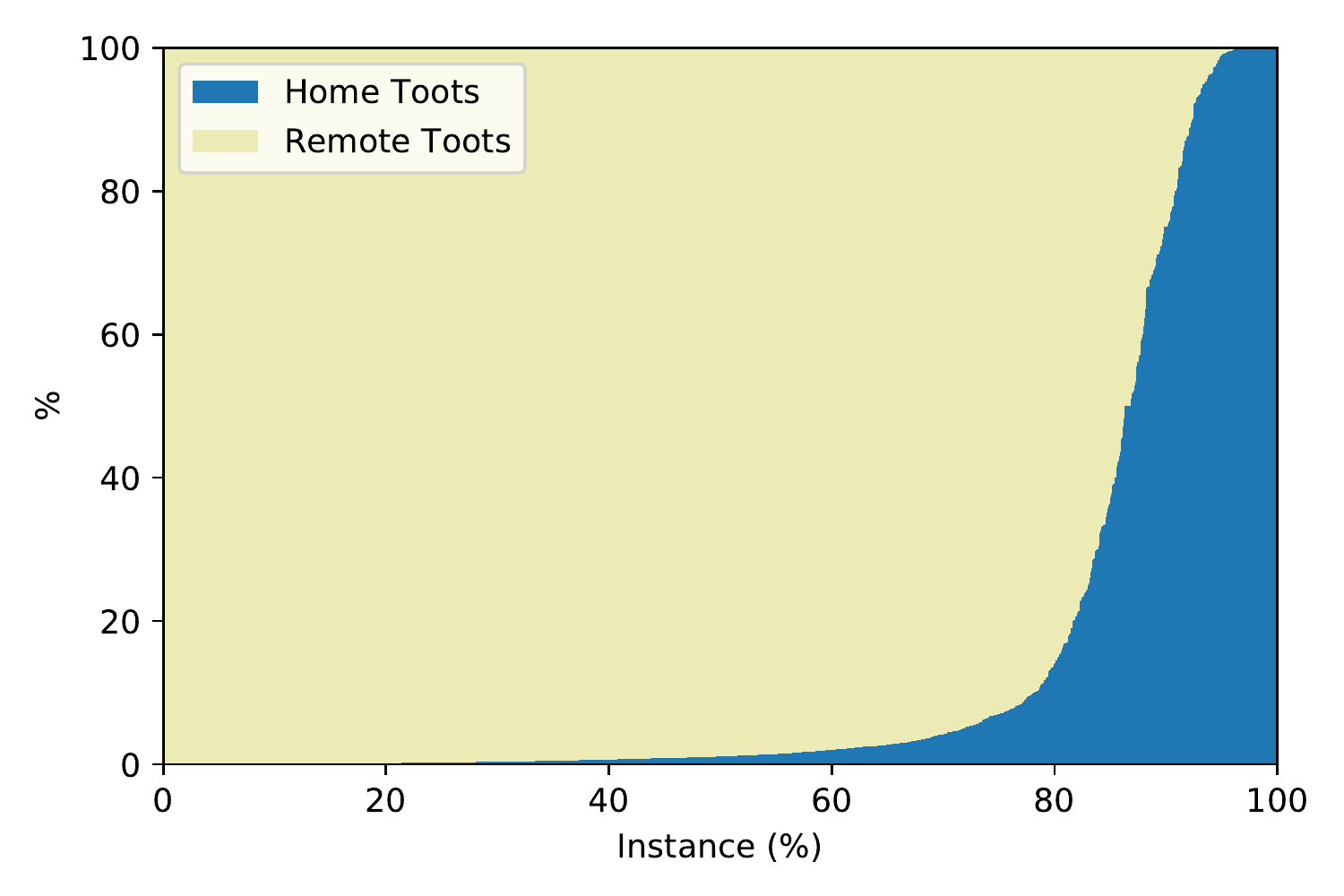}
	\vspace{-0.3cm}
	\caption{Ratio of home toots (produced on the instance) to remote toots (replicated from other ones).}%
	\label{fig:home_vs_away_toots}
\end{figure}

\begin{figure*}[t]
\centering
\includegraphics[width=0.865\textwidth]{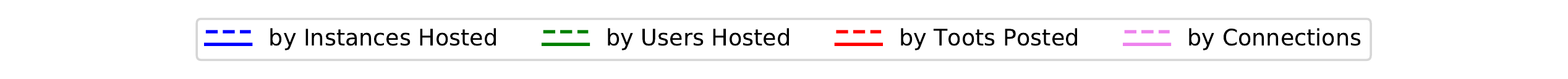}\\[-1ex]
\includegraphics[width=0.049\textwidth]{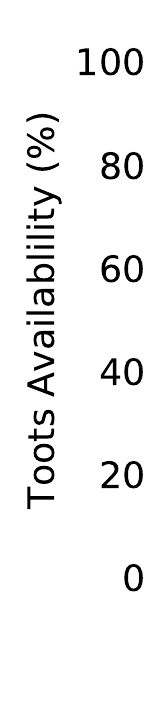}
 \subfigure[No replication]{\includegraphics[width=0.20\textwidth]{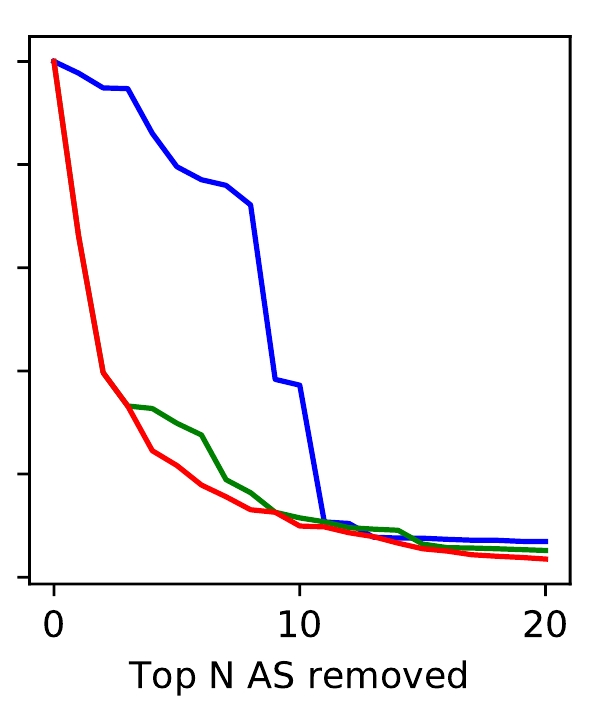}\label{fig:contentAvInsRemoval-as-nr}}
 \hspace{0.2cm}
 \subfigure[No replication]{\includegraphics[width=0.20\textwidth]{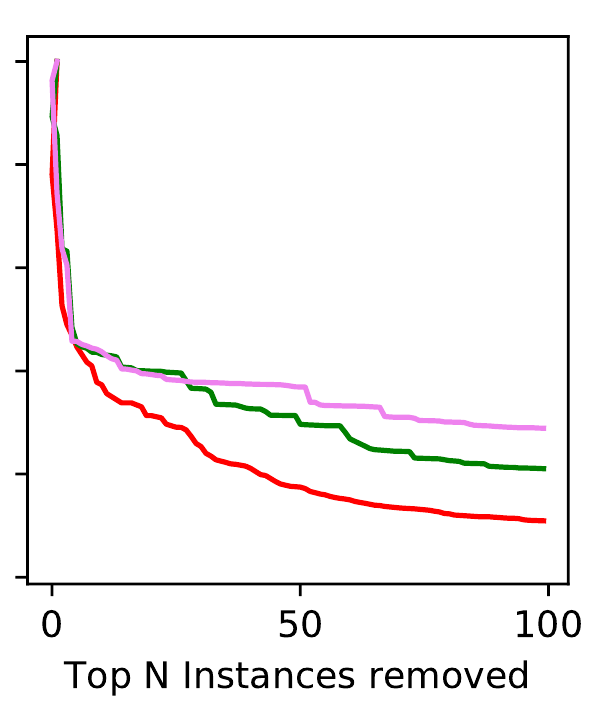}\label{fig:contentAvInsRemoval-ins-nr}}
 \hspace{0.2cm}
 \subfigure[Subscription Replication]{\includegraphics[width=0.20\textwidth]{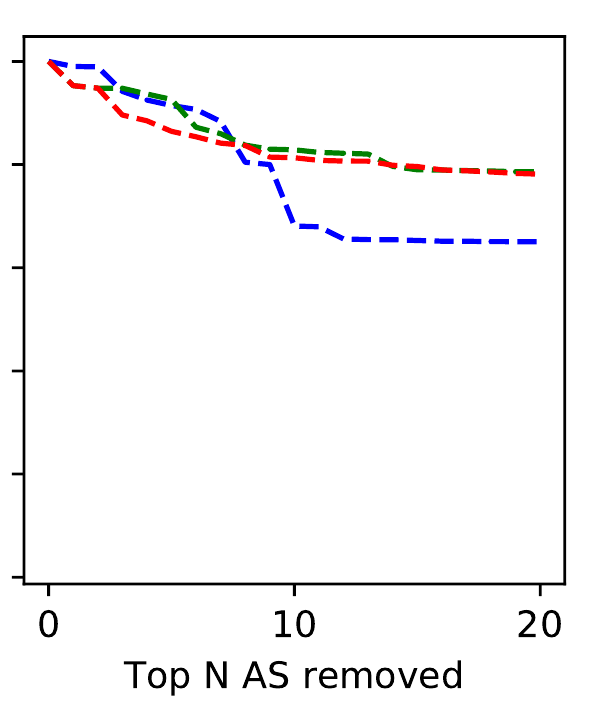}\label{fig:contentAvInsRemoval-as-r}}
 \hspace{0.2cm}
 \subfigure[Subscription Replication]{\includegraphics[width=0.20\textwidth]{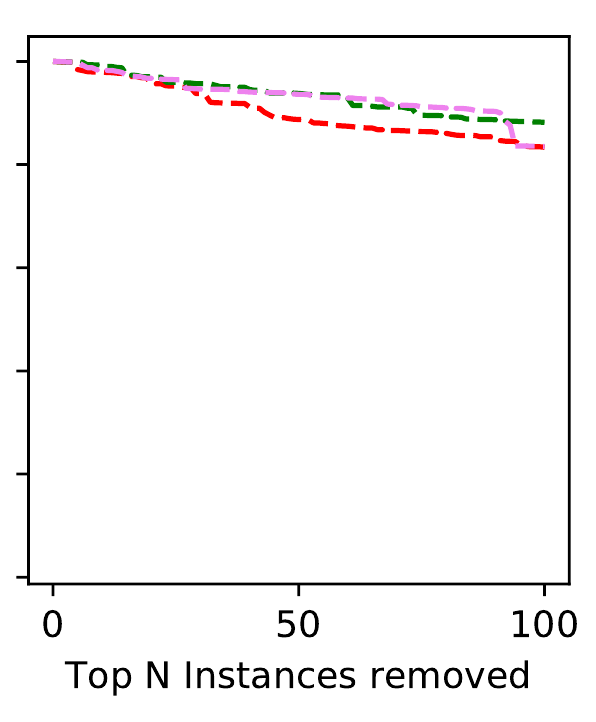}\label{fig:contentAvInsRemoval-ins-r}}
 \vspace{-0.2cm}
\caption{Availability of toots based on $(a)$ removing ASes, with ASes ranked based on \#instances, \#toots and \#users hosted; and $(b)$ removing instances, ranked by \#users, \#toots, and \#connections with other instances. In $(c)$ and $(d)$, we report the toot availability when replicating across all instances that follow them.}
\label{fig:contentAvInsRemoval}
\vspace*{-0.15cm}
\end{figure*}

\subsection{Breaking the Content Federation}  
\label{sec:content}
The above process of federation underpins the delivery of toots across the social graph.
For example, when a user shares a toot, it results in the toot being shared with the instances of all their followers.
Although we obviously cannot validate if a user reads a toot, we next explore the importance of federation for propagating content to timelines.

\paragraphbe{Role of Remote Toots.}
Aiming to measure the prevalence of federated remote toots, Figure~\ref{fig:home_vs_away_toots} plots the proportion of home \vs remote toots taken from the federated timeline (see Section~\ref{sec:primerMast}) of every instance in our dataset, ordered by the percentage of home toots.
The majority of toots on the federated timeline are actually generated by remote instances: 78\% of instances produce under 10\% of their own toots. In the most extreme case, we see that 5\% of instances are entirely reliant on remote toots, and generate no home toots themselves.
This suggests that some highly influential central instances operate as `feeders' to the rest of the network. 
Also, the more toots an instance generates, the higher the probability of them being replicated to other instances (correlation 0.97), thus highlighting the importance of a small number of feeders, without whom smaller instances would be unable to bootstrap. 
This is another inherent form of centralisation, which any social system will struggle to deviate from.

These results motivate us to measure the impact of instance and AS failures on toot availability. 
We evaluate three scenarios: 
\one~where a toot is exclusively hosted on its home instance, and
fetched by the remote instance on demand (denoted as ``no replication''); 
\two~where a toot is actively replicated to any instances with users that follow the toot's author (``subscription replication''); and 
\three~where a toot is replicated onto a random set of $n$ instances (``random replication'').
Mastodon partly supports option \two, but replicated toots are only temporarily cached and they are not globally indexed, \ie they are only visible to users local to the instance where the replica is. 
For these experiments, we assume a scenario where toots are globally indexed, \eg via a Distributed Hash Table~\cite{zhao2004tapestry}, allowing users to access replicated toots from any instance. For simplicity, we treat all toots as equal, even though in practice more recent toots are likely to be more important.

\paragraphbe{No replication.} In Figure~\ref{fig:contentAvInsRemoval-as-nr}, we measure the availability of toots when removing entire ASes; and in Figure~\ref{fig:contentAvInsRemoval-ins-nr}, we examine availability of toots when individual instances fail. Both plots depict results without replication enabled.
In both cases, toot availability declines rapidly in the face of failures. 
Removing the top 10 instances (ranked by number of toots) results in the deletion of 62.69\% toots from Mastodon.
Removing the top 10 ASes (again, ranked by number of toots) results in 90.1\% of toots becoming unavailable.
Therefore, we argue that running Mastodon without replication is not a viable option if resilience is to be a priority.

\paragraphbe{Subscription-based Replication.}
Figures~\ref{fig:contentAvInsRemoval-as-r} and~\ref{fig:contentAvInsRemoval-ins-r} report the availability of toots if they are replicated onto the instances that follow them, \ie via federated links. We consider any toot as available if there is at least one live Mastodon instance  holding a replica, and assume the presence of a global index (such as a Distributed Hash Table) to discover toots in such replicas. 

Availability improves using this method. 
For example, removing the top 10 instances now only results in 2.1\%  of toots becoming unavailable (as compared to 62.69\% without replication).
The equivalent values when removing the top 10 ASes are 18.66\% with replication (\vs 90.1\% without).

\begin{figure}[t]
	\centering
	\includegraphics[width=0.90\linewidth]{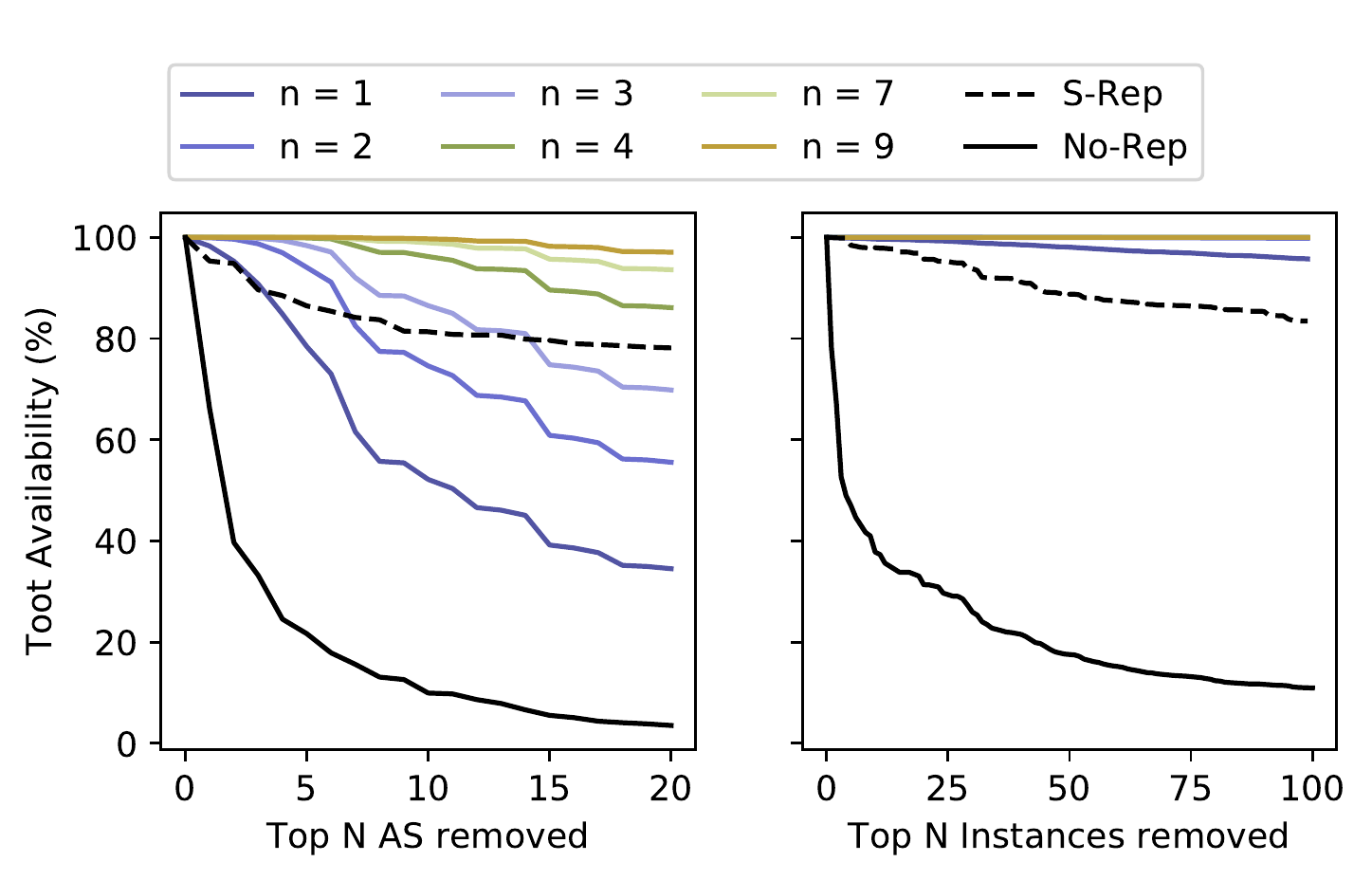}
	\vspace{-0.2cm}
    \caption{Availability of toots when removing top instances (measured by \#toots) with random replication. Lines for $n>4$ overlap indicating availability above 99\% (97\%) in case of instance (AS) failures.}
	\label{fig:instanceRApr50}
		\vspace{-0.3cm}
\end{figure}

\paragraphbe{Random Replication.} 
Above, we assumed that toots are only replicated to the followers' instances. We now experiment with a random replication strategy that copies each toot onto $n$ random instances. We test for $n=\{1, 2, 3, 4, 7, 9\}$, alongside no replication (No-rep) and subscription-based replication (S-Rep). We do this for all toots and present the results in Figure~\ref{fig:instanceRApr50}. In each iteration, we remove the current remaining top instance (as ranked by number of toots), and check the availability of toots according to the different replication strategies. %

The figures shows that random replication substantially outperforms subscription-based replication.
This is due to the biased way that subscription-based replication works, in which we find that 9.7\% of toots have no replication due to a lack of followers, yet 23\% of toots have more than 10 replicas because they are authored by popular users.
After removing 25 instances, subscription-based replication has 95\% availability, whereas 99.2\% availability could have been achieved with just 1 \emph{random} replication.
More prominently, subscription-based replication tends to place \emph{all} replicas onto a small set of popular instances (\ie where the followers are), due to the skewed popularity distribution of users.
This is yet another consequence of the natural centralisation that these systems experience. 
Thus, removing these popular instances will remove not only the original toot but also the replica(s).  

In contrast, the random strategy  distributes the load more evenly, such that instance failures impact fewer replicas of the toots. 
There are a number of practical concerns that will impact such strategies. Most notably, it would be important to weight replication based on the resources available at the instance (\eg storage).

\section{Related Work}
\label{sec:rw}

\paragraphbe{Decentralised Social Networks.}
Several efforts have worked toward building distributed social network platforms. 
At first, these were largely peer-to-peer, \eg LifeSocial.KOM~\cite{graffi2011lifesocial} and PeerSON~\cite{buchegger2009peerson}, and relied on entirely decentralised protocols. Unfortunately, they did not achieve widespread adoption partly due to challenges related to both performance~\cite{anjum2017survey} and reliability~\cite{kaune2010unraveling}.
Thus, a new wave of \DW platforms has emerged that rely on a server-based federated model, \eg Mastodon. %
Before that, the most successful platform was Diaspora, which saw growth around 2011~\cite{guidi2018managing}. %
Diaspora offers Facebook-like functionality using a federated model similar to Mastodon. However, unlike Mastodon, it relies on bespoke protocols (rather than standards), and its user population has since stagnated. 

Since Diaspora's creation, several (semi-)standard federation protocols have been proposed allowing instances to exchange data~\cite{silva2017implementing}. 
These include OStatus~\cite{ostatus}, which allows real-time interaction between instances; WebFinger~\cite{jones2013webfinger} for discovering information about entities on other instances; and ActivityPub~\cite{activitypub} for publishing information about social activities.
Attempts have also been  made to standardise the format of these data structures, \eg ActivityStreams~\cite{activitystream}. 
This standardisation efforts have had a dramatic impact on the \DW.
For example, both Mastodon and PeerTube use ActivityStreams and ActivityPub; thus, they can exchange data. An overview of these various protocol can be found in~\cite{guy2017social}.

Researchers have also looked at security and privacy in the \DW~\cite{taheri2015security,schwittmann2013sonet}, mostly around securing data management. %
For example, various projects have attempted to decentralise data, \eg DataBox~\cite{perera2017valorising}, SOLID~\cite{mansour2016demonstration}, and  SocialGate~\cite{koll2017socialgate}. 
These operate local datastores for individual users, \eg running on a physical home appliance. Applications wishing to access user data must be granted permission, potentially through a prior negotiation mechanisms. 

\paragraphbe{Social Network Measurements.}
A number of measurement studies have analysed ``centralised'' social networks like Facebook~\cite{ugander2011anatomy,raman2018facebook} and Twitter~\cite{kwak2010twitter,cha2010measuring,gilani2019large}.
These have revealed a range of properties, including attributes of the social graph and content generation. 
Bielenberg \etal performed the first study of a \DW application, Diaspora~\cite{bielenberg2012growth}.
When inspecting its growth, they found a network far smaller than the one we observe on Mastodon. %
There has been a small set of recent works that focus on Mastodon. Zignani \etal collected and released Mastodon datasets, as well as exploring several features, \eg the social graph, placement of instances and content warnings~\cite{zignani2018follow,zignani2019mastodon}. 
Also, studies have focused on friend recommendations~\cite{trienes2018recommending} and sentiment analysis~\cite{cerisara2018multi}.
We complement these works with a focus on availability, covering the key aspects of federation. We also inspect the nature and deployment of instances, as well as their topical interests. To the best of our knowledge, this paper constitutes the largest study to date of Mastodon.

\section{Conclusion}
\label{sec:conclusion}

This paper presented a large-scale measurement study of the Decentralised Web (\DW) through the lens of Mastodon. 
We focused on exploring challenges arising from two key innovations introduced by the \DW: \one~the decomposition of a global service into many independent \emph{instances}; and \two~the process of \emph{federation}, whereby these instances collaborate and interact. 

We found that Mastodon's design decision of giving everyone the ability to establish an independent instance of their own has led to an active ecosystem, with instances covering a wide variety of topics. However, a common theme in our work has been the discovery of apparent forms of centralisation within Mastodon. 
For example, 10\% of instances host almost half of the users, and certain categories exhibit remarkable reliance on a small set of instances. This extends to hosting practices, with three ASes hosting almost two third of the users. 

Our simulations further confirmed that these natural pressures towards centralisation lead to potential points of failure. 
For example, it was possible to reduce the LCC in the federation graph from 92\% of all users to just 46\% via the removal of five ASes. Similarly, outages in just 10 instances can remove almost half of all toots. This is not a theoretical problem: we discovered regular instance (and even occasional AS) outages.
Looking for possible mitigations, we experimented with simple replication strategies to find that availability can be dramatically improved by copying toots onto secondary instances, \ie by reducing the level of centralisation. Interestingly, the subscription-based strategy (loosely employed by Mastodon currently) is not as effective as a random strategy, due to the propensity to replicate toots onto the same set of instances where the followers are based. 

We argue that if these problems are ignored, the \DW may risk converging towards a semi-centralised system. As part of future work, we plan to explore the longer term properties of the \DW more generally. 
We will also work on mitigations to some of the identified concerns (beyond the toot replication discussed in Section~\ref{sec:content}), including decentralised defenses against, \eg malicious bots. One example of a possible mitigation is the existing instance blocking supported by Mastodon; our future work will investigate the impact that this has on the social graph and how it can be used to filter malicious content. Importantly, we argue that mitigations should not depend on the exposure of user information to remote 

As part of future work, we plan to provide a longitudinal analysis of Mastodon with respect to the issues identified in this paper as well as study the effect in terms of hate speech of alt-right communities like Gab~\cite{zannettou2018gab} starting to use Mastodon forks~\cite{gabmastodon}.

\subsection*{Acknowledgements}
We would like to thank Eliot Berriot for creating \url{mnm.social} as well as the ACM IMC Program Committee and in particular our shepherd, Christo Wilson, for their comments and feedback. This research was funded by The Alan Turing Institute's programme on Defence and Security, and supported by The Alan Turing Institute under the EPSRC grant EP/N510129/1 and the EU Commission under the H2020 ENCASE project (Grant No. 691025).

\small
\bibliographystyle{abbrv} %

\begin{thebibliography}{10}

\bibitem{activitypub}
{ActivityPub}.
\newblock \url{https://www.w3.org/TR/activitypub/}, 2018.

\bibitem{activitystream}
ActivityStream.
\newblock \url{http://www.w3.org/ns/activitystreams}, 2017.

\bibitem{albert2000error}
R.~Albert, H.~Jeong, and A.-L. Barab{\'a}si.
\newblock {Error and attack tolerance of complex networks}.
\newblock {\em Nature}, 406(6794), 2000.

\bibitem{anjum2017survey}
N.~Anjum, D.~Karamshuk, M.~Shikh-Bahaei, and N.~Sastry.
\newblock {Survey on peer-assisted content delivery networks}.
\newblock {\em Computer Networks}, 116, 2017.

\bibitem{bielenberg2012growth}
A.~Bielenberg, L.~Helm, A.~Gentilucci, D.~Stefanescu, and H.~Zhang.
\newblock {The growth of Diaspora -- A decentralized online social network in
  the wild}.
\newblock In {\em {INFOCOM Workshops}}, 2012.

\bibitem{braunstein2016network}
A.~Braunstein, L.~Dall'Asta, G.~Semerjian, and L.~Zdeborov{\'a}.
\newblock {Network dismantling}.
\newblock {\em Proceedings of the National Academy of Sciences}, 113(44), 2016.

\bibitem{buchegger2009peerson}
S.~Buchegger, D.~Schi{\"o}berg, L.-H. Vu, and A.~Datta.
\newblock {PeerSoN: P2P social networking: early experiences and insights}.
\newblock In {\em {EuroSys Workshop on Social Network Systems}}, 2009.

\bibitem{asRankCaida}
CAIDA.
\newblock {Ranking of Autonomous Systems}.
\newblock \url{http://as-rank.caida.org/}, 2019.

\bibitem{cerisara2018multi}
C.~Cerisara, S.~Jafaritazehjani, A.~Oluokun, and H.~Le.
\newblock {Multi-task dialog act and sentiment recognition on Mastodon}.
\newblock {\em arXiv:1807.05013}, 2018.

\bibitem{cha2010measuring}
M.~Cha, H.~Haddadi, F.~Benevenuto, and P.~K. Gummadi.
\newblock {Measuring user influence in twitter: The million follower fallacy}.
\newblock In {\em {ICWSM}}, 2010.

\bibitem{crtsh}
Comodo.
\newblock {Comodo Launches New Digital Certificate Searchable Web Site}.
\newblock \url{https://bit.ly/2k27p64}, 2015.

\bibitem{doerr2012rumors}
B.~Doerr, M.~Fouz, and T.~Friedrich.
\newblock {Why rumors spread fast in social networks}.
\newblock {\em Communications of the ACM}, 55(6), 2012.

\bibitem{verge17}
M.~Farokhmanesh.
\newblock {A beginner's guide to Mastodon, the hot new open-source Twitter
  clone}.
\newblock
  \url{https://www.theverge.com/2017/4/7/15183128/mastodon-open-source-twitter-clone-how-to-use},
  2017.

\bibitem{fedinfo}
T.~Federation.
\newblock \url{https://the-federation.info/}, 2019.

\bibitem{gilani2019large}
Z.~Gilani, R.~Farahbakhsh, G.~Tyson, and J.~Crowcroft.
\newblock {A Large-scale Behavioural Analysis of Bots and Humans on Twitter}.
\newblock {\em ACM Transactions on the Web (TWEB)}, 13(1), 2019.

\bibitem{gilani2017bots}
Z.~Gilani, R.~Farahbakhsh, G.~Tyson, L.~Wang, and J.~Crowcroft.
\newblock {Of Bots and Humans (on Twitter)}.
\newblock In {\em {ASONAM}}, 2017.

\bibitem{giotsas2017detecting}
V.~Giotsas, C.~Dietzel, G.~Smaragdakis, A.~Feldmann, A.~Berger, and E.~Aben.
\newblock {Detecting peering infrastructure outages in the wild}.
\newblock In {\em {ACM SIGCOMM}}, 2017.

\bibitem{graffi2011lifesocial}
K.~Graffi, C.~Gross, D.~Stingl, D.~Hartung, A.~Kovacevic, and R.~Steinmetz.
\newblock {LifeSocial. KOM: A secure and P2P-based solution for online social
  networks}.
\newblock In {\em {CCNC}}, 2011.

\bibitem{guidi2018managing}
B.~Guidi, M.~Conti, A.~Passarella, and L.~Ricci.
\newblock {Managing social contents in Decentralized Online Social Networks: A
  survey}.
\newblock {\em Online Social Networks and Media}, 7, 2018.

\bibitem{gunawi2016does}
H.~S. Gunawi, M.~Hao, R.~O. Suminto, A.~Laksono, A.~D. Satria, J.~Adityatama,
  and K.~J. Eliazar.
\newblock {Why does the cloud stop computing?: Lessons from hundreds of service
  outages}.
\newblock In {\em {ACM SoCC}}, 2016.

\bibitem{guy2017social}
A.~Guy.
\newblock {Social web protocols}.
\newblock W3C Technical Report, 2017.

\bibitem{archive}
{Internet Archive}.
\newblock {Twitter Outages}.
\newblock
  \url{https://web.archive.org/web/20110828003545/http://stats.pingdom.com/wx4vra365911/23773/2007/02},
  2007.

\bibitem{jmir2018}
S.~Joglekar, N.~Sastry, N.~S. Coulson, S.~J. Taylor, A.~Patel, R.~Duschinsky,
  A.~Anand, M.~Jameson~Evans, C.~J. Griffiths, A.~Sheikh, P.~Panzarasa, and
  A.~De~Simoni.
\newblock {How Online Communities of People With Long-Term Conditions Function
  and Evolve: Network Analysis of the Structure and Dynamics of the Asthma UK
  and British Lung Foundation Online Communities}.
\newblock {\em J Med Internet Res}, 20(7), 2018.

\bibitem{jones2013webfinger}
P.~Jones, G.~Salgueiro, M.~Jones, and J.~Smarr.
\newblock {WebFinger}.
\newblock \url{https://tools.ietf.org/html/rfc7033}, 2013.

\bibitem{kaune2010unraveling}
S.~Kaune, R.~C. Rumin, G.~Tyson, A.~Mauthe, C.~Guerrero, and R.~Steinmetz.
\newblock {Unraveling BitTorrent's file unavailability: Measurements and
  analysis}.
\newblock In {\em {P2P}}, 2010.

\bibitem{koll2017socialgate}
D.~Koll, D.~Lechler, and X.~Fu.
\newblock {SocialGate: Managing large-scale social data on home gateways}.
\newblock In {\em {IEEE ICNP}}, 2017.

\bibitem{kwak2010twitter}
H.~Kwak, C.~Lee, H.~Park, and S.~Moon.
\newblock {What is Twitter, a social network or a news media?}
\newblock In {\em {WWW}}, 2010.

\bibitem{failWhale}
A.~Lafrance.
\newblock {The Story of Twitter's Fail Whale}.
\newblock
  \url{https://www.theatlantic.com/technology/archive/2015/01/the-story-behind-twitters-fail-whale/384313/},
  2015.

\bibitem{lakhina2004characterization}
A.~Lakhina, M.~Crovella, and C.~Diot.
\newblock {Characterization of network-wide anomalies in traffic flows}.
\newblock In {\em {ACM SIGCOMM}}, 2004.

\bibitem{leskovec2012learning}
J.~Leskovec and J.~J. Mcauley.
\newblock {Learning to discover social circles in ego networks}.
\newblock In {\em {NIPS}}, 2012.

\bibitem{letsencrypt}
{Let's Encrypt - FAQ}.
\newblock \url{https://letsencrypt.org/docs/faq/}, 2017.

\bibitem{mansour2016demonstration}
E.~Mansour, A.~V. Sambra, S.~Hawke, M.~Zereba, S.~Capadisli, A.~Ghanem,
  A.~Aboulnaga, and T.~Berners-Lee.
\newblock {A demonstration of the solid platform for social web applications}.
\newblock In {\em {WWW}}, 2016.

\bibitem{mastodon-git}
{Mastodon}.
\newblock \url{https://github.com/tootsuite/mastodon}, 2018.

\bibitem{gabmastodon}
Mastodon.
\newblock {Statement on Gab's fork of Mastodon}.
\newblock
  \url{https://blog.joinmastodon.org/2019/07/statement-on-gabs-fork-of-mastodon/},
  2019.

\bibitem{fediverse}
F.~network.
\newblock \url{https://fediverse.network/reports/2018}, 2018.

\bibitem{wired2}
B.~Nystedt.
\newblock {Tired of Twitter? Join me on Mastodon}.
\newblock \url{https://www.wired.com/story/join-mastodon-twitter-alternative/},
  2018.

\bibitem{ostatus}
{Ostaus 1.0 Protocol Specification}.
\newblock
  \url{https://www.w3.org/community/ostatus/wiki/images/9/93/OStatus_1.0_Draft_2.pdf},
  2010.

\bibitem{perera2017valorising}
C.~Perera, S.~Y. Wakenshaw, T.~Baarslag, H.~Haddadi, A.~K. Bandara, R.~Mortier,
  A.~Crabtree, I.~C. Ng, D.~McAuley, and J.~Crowcroft.
\newblock {Valorising the IoT databox: creating value for everyone}.
\newblock {\em Transactions on Emerging Telecommunications Technologies},
  28(1), 2017.

\bibitem{raman2018facebook}
A.~Raman, G.~Tyson, and N.~Sastry.
\newblock {Facebook (A) Live? Are live social broadcasts really broadcasts?}
\newblock {\em The Web Conference}, 2018.

\bibitem{schwittmann2013sonet}
L.~Schwittmann, C.~Boelmann, M.~Wander, and T.~Weis.
\newblock {SoNet--Privacy and Replication in Federated Online Social Networks}.
\newblock In {\em {Distributed Computing Systems Workshops}}, 2013.

\bibitem{silva2017implementing}
G.~Silva, L.~Reis, A.~Terceiro, P.~Meirelles, and F.~Kon.
\newblock {Implementing Federated Social Networking: Report from the Trenches}.
\newblock In {\em {OpenSym}}, 2017.

\bibitem{what_is_mastodon}
C.~Steele.
\newblock {What Is Mastodon and Will It Kill Twitter?}
\newblock
  \url{https://au.pcmag.com/social-networking-1/47343/what-is-mastodon-and-will-it-kill-twitter},
  2017.

\bibitem{stigler1958economies}
G.~J. Stigler.
\newblock {The economies of scale}.
\newblock {\em The Journal of Law and Economics}, 1, 1958.

\bibitem{taheri2015security}
S.~Taheri-Boshrooyeh, A.~K{\"u}p{\c{c}}{\"u}, and {\"O}.~{\"O}zkasap.
\newblock {Security and privacy of distributed online social networks}.
\newblock In {\em {Distributed Computing Systems Workshops}}, 2015.

\bibitem{fortune18}
H.~Timms and J.~Heimans.
\newblock {Commentary: \#DeleteFacebook Is Just the Beginning. Here's the
  Movement We Could See Next}.
\newblock
  \url{http://fortune.com/2018/04/16/delete-facebook-data-privacy-movement/},
  2018.

\bibitem{torok2017cascading}
J.~T{\"o}r{\"o}k and J.~Kert{\'e}sz.
\newblock {Cascading collapse of online social networks}.
\newblock {\em Scientific reports}, 7(1), 2017.

\bibitem{trienes2018recommending}
J.~Trienes, A.~T. Cano, and D.~Hiemstra.
\newblock {Recommending Users: Whom to Follow on Federated Social Networks}.
\newblock {\em arXiv:1811.09292}, 2018.

\bibitem{tyson2013demystifying}
G.~Tyson, Y.~Elkhatib, N.~Sastry, and S.~Uhlig.
\newblock {Demystifying porn 2.0: A look into a major adult video streaming
  website}.
\newblock In {\em {ACM IMC}}, 2013.

\bibitem{ugander2011anatomy}
J.~Ugander, B.~Karrer, L.~Backstrom, and C.~Marlow.
\newblock {The anatomy of the facebook social graph}.
\newblock {\em arXiv:1111.4503}, 2011.

\bibitem{wilson2009user}
C.~Wilson, B.~Boe, A.~Sala, K.~P. Puttaswamy, and B.~Y. Zhao.
\newblock {User interactions in social networks and their implications}.
\newblock In {\em {ACM EuroSys}}, 2009.

\bibitem{yagan2013conjoining}
O.~Yagan, D.~Qian, J.~Zhang, and D.~Cochran.
\newblock {Conjoining speeds up information diffusion in overlaying
  social-physical networks}.
\newblock {\em IEEE Journal on Selected Areas in Communications}, 31(6), 2013.

\bibitem{zannettou2018gab}
S.~Zannettou, B.~Bradlyn, E.~De~Cristofaro, H.~Kwak, M.~Sirivianos,
  G.~Stringini, and J.~Blackburn.
\newblock {What is Gab: A bastion of free speech or an alt-right echo chamber}.
\newblock In {\em WWW Companion}, 2018.

\bibitem{zhao2004tapestry}
B.~Y. Zhao, L.~Huang, J.~Stribling, S.~C. Rhea, A.~D. Joseph, and J.~D.
  Kubiatowicz.
\newblock {Tapestry: A resilient global-scale overlay for service deployment}.
\newblock {\em IEEE Journal on Selected Areas in Communications}, 22(1), 2004.

\bibitem{zignani2018follow}
M.~Zignani, S.~Gaito, and G.~P. Rossi.
\newblock {Follow the ``Mastodon'': Structure and Evolution of a Decentralized
  Online Social Network}.
\newblock In {\em {ICWSM}}, 2018.

\bibitem{zignani2019mastodon}
M.~Zignani, C.~Quadri, A.~Galdeman, S.~Gaito, and G.~P. Rossi.
\newblock {Mastodon Content Warnings: Inappropriate Contents in a Microblogging
  Platform}.
\newblock In {\em {ICWSM}}, 2019.

\end{thebibliography}

\end{document}